\documentclass[preprint2]{aastex61}

\submitjournal{Astrophysical Journal}
\accepted{August 7, 2017}

\usepackage{natbib}
\newcommand{\stsp}{\texttt{STSP}}
\newcommand{\kepler}{\textit{Kepler}}
\usepackage{amsmath}
\usepackage{color}

\shorttitle{HAT-P-11: Evidence for a Solar-like Dynamo}
\shortauthors{Morris et al.}
\submitjournal{ApJ}

\begin{document}

\title{The Starspots of HAT-P-11: Evidence for a solar-like dynamo}

\author{Brett M. Morris}
\affiliation{Astronomy Department, University of Washington, Seattle, WA 98119, USA}

\author{Leslie Hebb}
\affiliation{Physics Department, Hobart and William Smith Colleges, Geneva, NY 14456, USA}

\author{James R. A. Davenport}
\affiliation{Department of Physics \& Astronomy, Western Washington University, Bellingham, WA 98225, USA}
\affiliation{NSF Astronomy and Astrophysics Postdoctoral Fellow}

\author{Graeme Rohn}
\affiliation{Physics Department, State University of New York at Cortland, Cortland, New York 13045-0900, USA }

\author{Suzanne L. Hawley}
\affiliation{Astronomy Department, University of Washington, Seattle, WA 98119, USA}

\email{bmmorris@uw.edu}

\begin{abstract}
We measure the starspot radii and latitude distribution on the K4 dwarf HAT-P-11 from \kepler\ short-cadence photometry. We take advantage of starspot occultations by its highly-misaligned planet to compare the spot size and latitude distributions to those of sunspots. We find that the spots of HAT-P-11 are distributed in latitude much like sunspots near solar activity maximum, with mean spot latitude of $\approx 16 \pm 1^\circ$. The majority of starspots of HAT-P-11 have physical sizes that closely resemble the sizes of sunspots at solar maximum. We estimate the mean spotted area coverage on HAT-P-11 is $3^{+6}_{-1}\%$, roughly two orders of magnitude greater than the typical solar spotted area.
\end{abstract}

%\object{HAT-P-11}

\keywords{starspots, sunspots, stellar activity, stellar activity cycles, Kepler photometry, stellar dynamo}

\section{Introduction}

The Sun is our local laboratory for understanding stellar magnetic activity. Centuries of sunspot observations and recent helioseismology results point towards the $\alpha\Omega$ dynamo mechanism as the source of solar magnetic activity. Solar magnetic fields are stored and amplified in poloidal and toroidal components, in the tachocline beneath the convective zone, until magnetic buoyancy causes them to rise. The buoyant magnetic flux tubes become visible as sunspots where they intersect with the photosphere \citetext{\citealp{Parker1955a, Parker1955b, Babcock1961}; \citealp[see reviews by][]{Charbonneau2010, Cheung2014, Hathaway2015}}. 

Magnetic activity on slowly rotating Sun-like stars is difficult to measure \citep{Saar1990} because the Sun has dark spots spanning only $\lesssim 0.5\%$ of its surface area at its most active, while spot areas of at least $\gtrsim 10\%$ are required to detect high S/N molecular absorption or Zeeman splitting. Polarization can characterize spots on resolved stars such as the Sun, but the opposite polarities in bipolar magnetic regions cancel one another in unresolved spot pairs, yielding little net polarization. As a result, most of our measurements of stellar activity come from stars much more active than the Sun \citep[see reviews by][]{Berdyugina2005, Reiners2012}.

Initial observations of a small sample of Sun-like stars show that the fraction of magnetic energy stored in the toroidal field decreases as rotation period increases \citep{Petit2008}. Spot temperatures and area covering fractions have been inferred from molecular absorption by TiO and OH in cool starspots of Sun-like stars \citep{Neff1995, ONeal1996, ONeal2001, ONeal2004}.  The properties of sunspots that are most informative for constraining dynamo theory, such as the physical sizes and latitude distributions of spots, are typically highly degenerate with these observing techniques.

Transiting exoplanets enable measurements of spot sizes and positions on their host stars. During an exoplanet transit, the flux lost at any instant is proportional to the intensity of the occulted portion of the stellar surface. Occultations of starspots by exoplanets are observed as positive flux anomalies in transit light curves, which are resolved in time by \kepler\ short-cadence photometry. \cite{Hebb2017} develop a photometric model for spotted stars which computes the observed light curve for spots of a given size and position on the stellar surface. \stsp\ simulates times during transit -- when the planet may or may not be occulting a spot -- and during the rest of the planetary orbit when the stellar rotation drives photometric variability.

In this work, we will make a direct comparison between the Sun and an exoplanet host star using the occultation mapping method. We measure starspot positions and sizes using the photometric model \stsp\ developed in \citet{Hebb2017}. \stsp\ simulates photometric time series measurements for stars with spotted surfaces and transiting planets. If we know the stellar orientation relative to the planet's orbit, the timing and morphology of spot occultations can be transformed into precise positions of starspots with the forward-modeling approach of \stsp.

\begin{figure}
\centering
\includegraphics[scale=0.4]{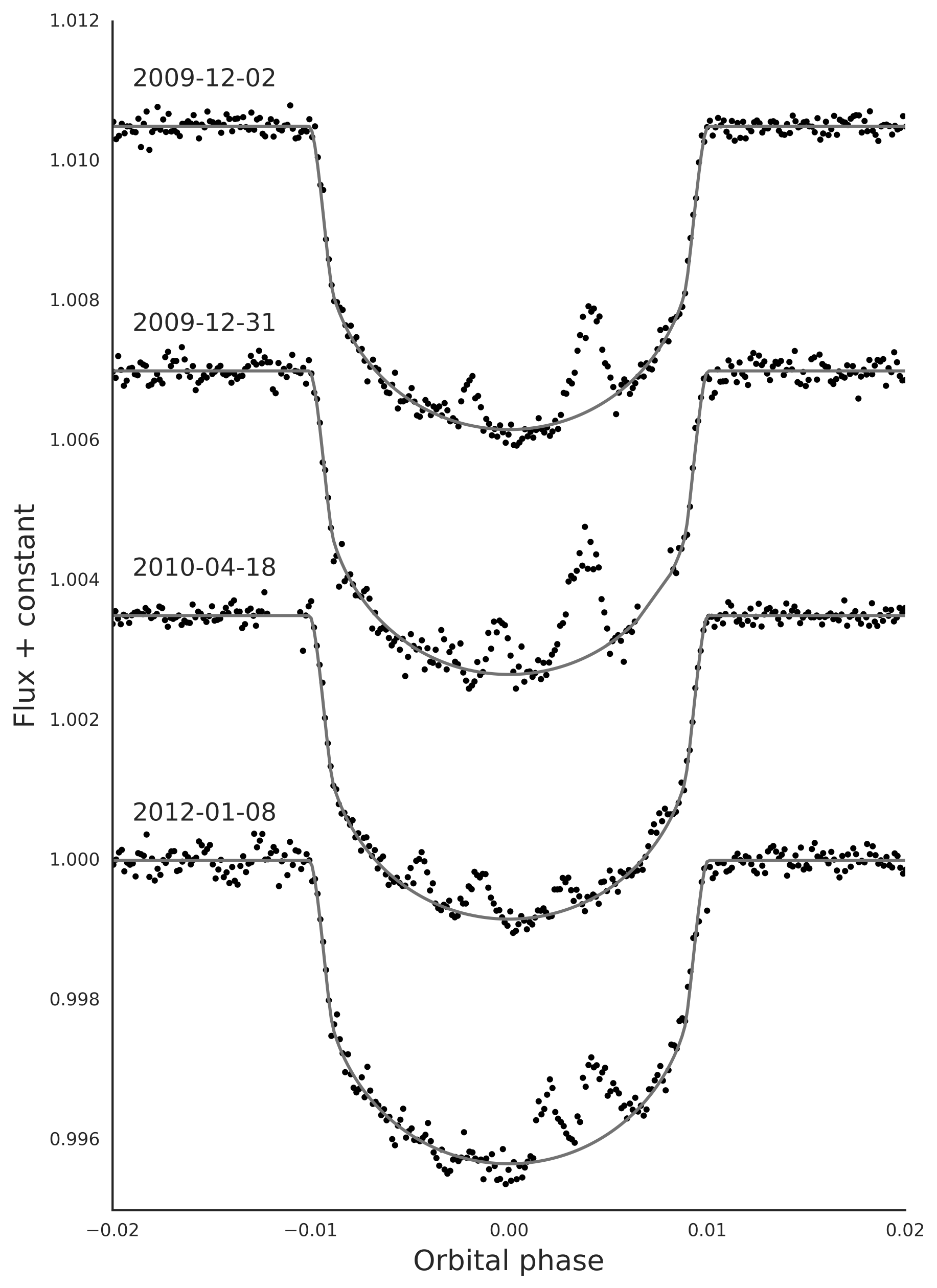}
\caption{Typical transit light curves of HAT-P-11 b. The points are \kepler\ fluxes, the curves are the best-fit transit model \citep{Mandel2002}. The positive anomalies during transit are occultations of starspots by the planet.}
\label{fig:transit_gallery}
\end{figure}

The active, transiting planet host star HAT-P-11 produces many spot occultations in its \kepler\ light curve -- see Figure~\ref{fig:transit_gallery} for examples. It is a K4 dwarf with a hot Neptune planet with orbital period $P = 4.88$ d, mass $M_p = 0.08 M_J$, and radius $R_p = 0.4 R_J$ \citep{bakos2010}. Observations of the Rossiter-McLaughlin effect revealed that HAT-P-11 b likely orbits over the poles of its host star \citep{Winn2010, Hirano2011, Sanchis-Ojeda2011}. Therefore the transit chord of the planet sweeps a path across one stellar longitude over many latitudes. The stellar rotation period and the orbital period of the planet are nearly commensurate at 6:1, which causes transits to occur near the same six stellar longitudes \citep{Beky2014a}.

Spot crossings of HAT-P-11 in the \kepler\ observations can be used to search for active stellar latitudes. \citet{Sanchis-Ojeda2011} and \citet{Deming2011} noted that the first few quarters of \kepler\ observations show spot occultations predominantly at two orbital phases, which they attribute to starspots which are concentrated into two active latitudes. \cite{Beky2014b} constructed a spot occultation model which they applied to HAT-P-11, and they estimate spot contrasts and sizes.

We introduce the \stsp\ model and solve for the inputs it requires in Section~\ref{sec:inputs}, and measure spot positions and sizes in Section~\ref{sec:stsp}. We compare the spot sizes, active latitudes, and spotted area coverage of HAT-P-11 to the Sun in Section~\ref{sec:results}. Finally, we discuss the properties of HAT-P-11's activity in Section~\ref{sec:conclusion}. 

\section{Starspot Modeling: Inputs for \stsp} \label{sec:inputs}

\citet{Hebb2017} developed a flux model for spotted stars with transiting planets called \stsp, which leverages spot occultations during planetary transits to break the spot position degeneracies. They illustrated the mapping technique on the young solar-like star Kepler-17. The alignment of the stellar spin and planetary orbit in that system confine the spot occultation observations to one narrow band of stellar latitudes. This alignment allowed them to probe the time evolution of spots, since the same spot was occulted multiple times in consecutive transits. In the HAT-P-11 system, the near-perpendicular misalignment of the stellar spin and planetary orbit alternatively allows us to probe spot positions as a function of latitude.

We first need to determine several input parameters that will be fixed in the \stsp\ flux model, enabling us to solve for the spot properties. In Section~\ref{sec:transit}, we fit for the orbital properties of the planet from the transit light curves. We solve for the initial spot positions with a simplified spot model in Sections~\ref{sec:spotoccmodel}-\ref{sec:friedrich}, which enables us to measure the approximate stellar inclination in Section~\ref{sec:i_s}, re-evaluate the spin-orbit obliquity in Section~\ref{sec:obliquity}, and to test our assumptions about spot contrasts in Section~\ref{sec:contrast}. With the approximate starspot positions derived from the simplified model fits, we explore the spot latitude-longitude-radius parameter space with the full \stsp\ forward model in Section~\ref{sec:stsp}. 

\stsp\ is a pure \texttt{C} code for calculating the variations in flux of a star due to spots, both in- and out-of-transit  \citep{Hebb2017}. We use \stsp\ because its prescription for the shapes of spot occultations are more realistic than the simple model in Sections~\ref{sec:spotoccmodel}-\ref{sec:friedrich}, and the correlations between MCMC parameters allow us to properly explore the degeneracies between starspot positions and sizes. \stsp\ can also solve for the properties of spots driving out-of-transit flux modulations, however in this work we consider only the spots detected in-transit, since the spot occultations yield tighter constraints on the spot properties than the out-of-transit flux variations.

\subsection{Orbital Properties of HAT-P-11 \lowercase{b}} \label{sec:transit}

To study the signal imparted by starspots on the transit light curve residuals, we must first remove the transit of HAT-P-11b from each light curve. It is non-trivial to derive the transit parameters for HAT-P-11 b since nearly all of the transits appear to be affected by starspots to some extent. We acknowledge that the most robust measurement of the transit properties would be obtained by fitting the light curve simultaneously for the transit and the occulted starspots, but the number of parameters in that fit is prohibitively large. Therefore, we opt to fit for the transit parameters on a subset of transits with minimal starspot anomalies, and to fix those transit parameters later when we fit for the starspot properties. In the next two sections, we outline the procedure for finding the orbital properties of HAT-P-11 b in spite of the abundant starspots.

\subsubsection{Light Curve Normalization} \label{sec:norm}

For the transit depths to be consistent in each transit light curve, an appropriate normalization for each transit must be chosen. Transit light curves are often normalized by the flux immediately before ingress and after egress. However, each transit light curve will have different relative depths if the total flux of the star is varying due to unocculted starspots \citep{Czesla2009, Carter2011, Csizmadia2013}. For example, if unocculted starspots dim the host star's flux by a factor $0 < \epsilon < 1$, the flux lost during transit $\Delta F$ is unaffected, but the total flux $\epsilon F$ is smaller, so the relative depth $\delta = \Delta F / (\epsilon F)$ is larger for the spotted star than for the unspotted star. The transits of HAT-P-11 likley have many occulted and unocculted starspots, and the transit depths would vary in time if simple out-of-transit flux normalization was used. In this section, we outline a normalization procedure that ideally yields transits of constant depth for stars with unocculted starspots, so that we can use a single depth parameter for all transits which corresponds to the square of the ratio of radii, $\delta \propto (R_p/R_s)^2$.

We assume that the peak flux of HAT-P-11 over a few stellar rotations is close to the unobscured brightness of the unspotted star\footnote{This assumption is revisited in the discussion in Section~\ref{sec:spotted_area}}. We then normalize all transit fluxes by: (1) fitting and subtracting a second-order polynomial to the out-of-transit Simple Aperture Photometry (SAP) fluxes near each transit; (2) adding the peak quarterly flux to each polynomial detrended transit; and (3) dividing each transit by the peak flux of each quarter. The subtraction by a second-order polynomial removes trends in flux due to stellar rotation, and the addition and division by the peak flux normalizes the out-of-transit fluxes to near-unity, while keeping the transit depths consistent between transits \citep{Hebb2017}. We must use the SAP flux because it is the unnormalized flux in units of electrons per second, rather than the PDCSAP flux which is already normalized.

\subsubsection{``Spotless'' Transits} \label{sec:spotless}

\begin{table}
\begin{center}
\begin{tabular}{lr}
Parameter & Measurement\\ \hline \hline
Orbital period [d] & $4.88780258 \pm 0.00000017$ \\
Mid-transit [JD] & $2454605.89146 \pm 0.000020$ \\
Depth $\approx \left(\frac{R_p}{R_*}\right)^2$ & $0.00340 \pm 0.00002$\\
Duration, $T_{14}$ [d] & $0.0980 \pm 0.0001$\\
$b$ & $0.141^{+0.052}_{-0.080}$ \\
$q_1 = (u_1 + u_2)^2$ & $0.48 \pm 0.01$ \\ 
$q_2 = \frac{u_1 }{2(u_1 + u_2)}$ & $0.46 \pm 0.01$ 
\end{tabular}
\end{center}
\caption{Transit light curve parameters for HAT-P-11 from the ten transits in Figure~\ref{fig:spotlesstransits}. $T_{14}$ is the duration between first and fourth contact; $q_1$ and $q_2$ are the limb-darkening parameters of \citet{Kipping2013}; $u_1$ and $u_2$ are the standard quadratic limb-darkening parameters. These parameters are fixed in the starspot fits.}
\label{tab:transitprops}
\end{table}

\begin{figure}
\centering
\includegraphics[scale=0.7]{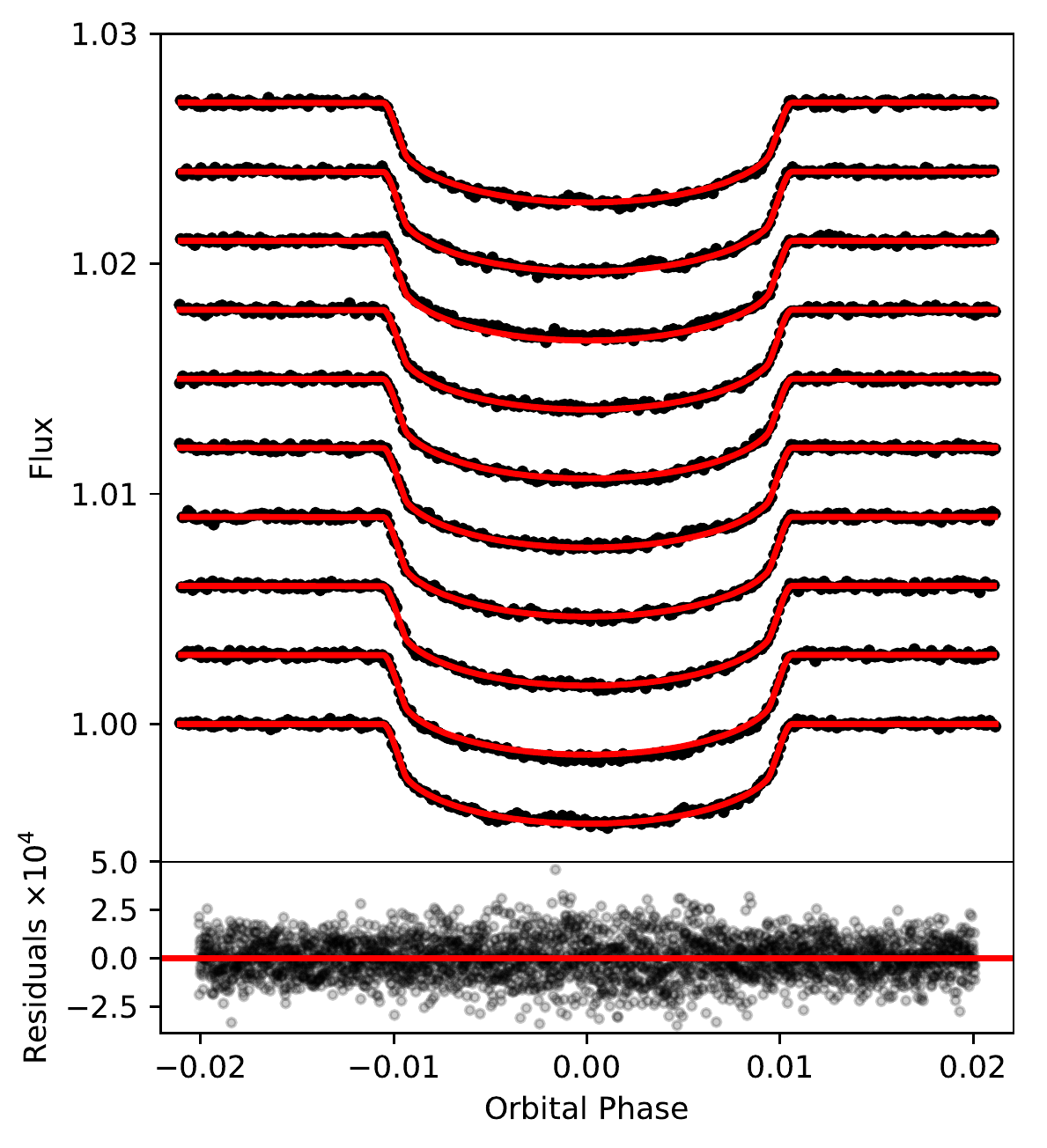}
\caption{The ten transits least affected by starspot crossings, which we fit for the planet orbital parameters listed in Table~\ref{tab:transitprops}.}
\label{fig:spotlesstransits}
\end{figure}

We select the ten transits with the fewest measurable starspot crossings to fit for the transit parameters. To identify the transits least perturbed by starspots, we fit a \citet{Mandel2002} transit light curve to each of the 205 normalized short-cadence transits in the full \kepler\ light curve. We hold the light curve parameters fixed, except for the depth which is allowed to vary, and optimize the light curve parameters using Levenburg-Marquardt least-squares minimization. We allow depth to vary because fits to transits with spot occultations (positive flux anomalies) will be biased towards smaller transit depths and higher $\chi^2$. We then select the ten transits with the smallest $\chi^2$. There are no significant starspot crossings visible by eye after this selection process, see Figure~\ref{fig:spotlesstransits}. 

We fit the ten transits for the orbital parameters and the stellar limb-darkening coefficients. We compute transit light curves with the \texttt{batman} package \citep{Kreidberg2015}, and sample the parameter posterior distributions with the affine-invariant Markov Chain Monte Carlo (MCMC) package \texttt{emcee} \citep{Foreman-Mackey2013}. The best-fit transit parameters are listed in Table~\ref{tab:transitprops}. 

The maximum likelihood planet-to-star radius ratio is $R_p/R_\star = 0.058 \pm 0.004$. This measurement is in agreement with \citet{Deming2011} ($0.0589 \pm 0.0002$), \citet{Southworth2011} ($0.058 \pm 0.001$), and \citet{bakos2010} ($0.0576 \pm 0.0009$).

Measurements of the mean stellar density of HAT-P-11 via asteroseismology and transit light curves have been reported by \citet{Christensen-Dalsgaard2010} and \citet{Southworth2011}, respectively. Using our transit parameters for HAT-P-11 b, we constrain the mean stellar density $\rho_s = 1.81 \pm 0.04 \rho_\odot$, which is similar to the preliminary asteroseismic measurement of $\rho_s = 1.7846 \pm 0.0006 \rho_\odot$, and smaller than the previous photometric measurement, $2.415 \pm 0.097\rho_\odot$.

\subsection{Spot Position Initial Conditions} \label{sec:spotoccmodel}

\begin{figure}
\centering
\includegraphics[scale=0.45]{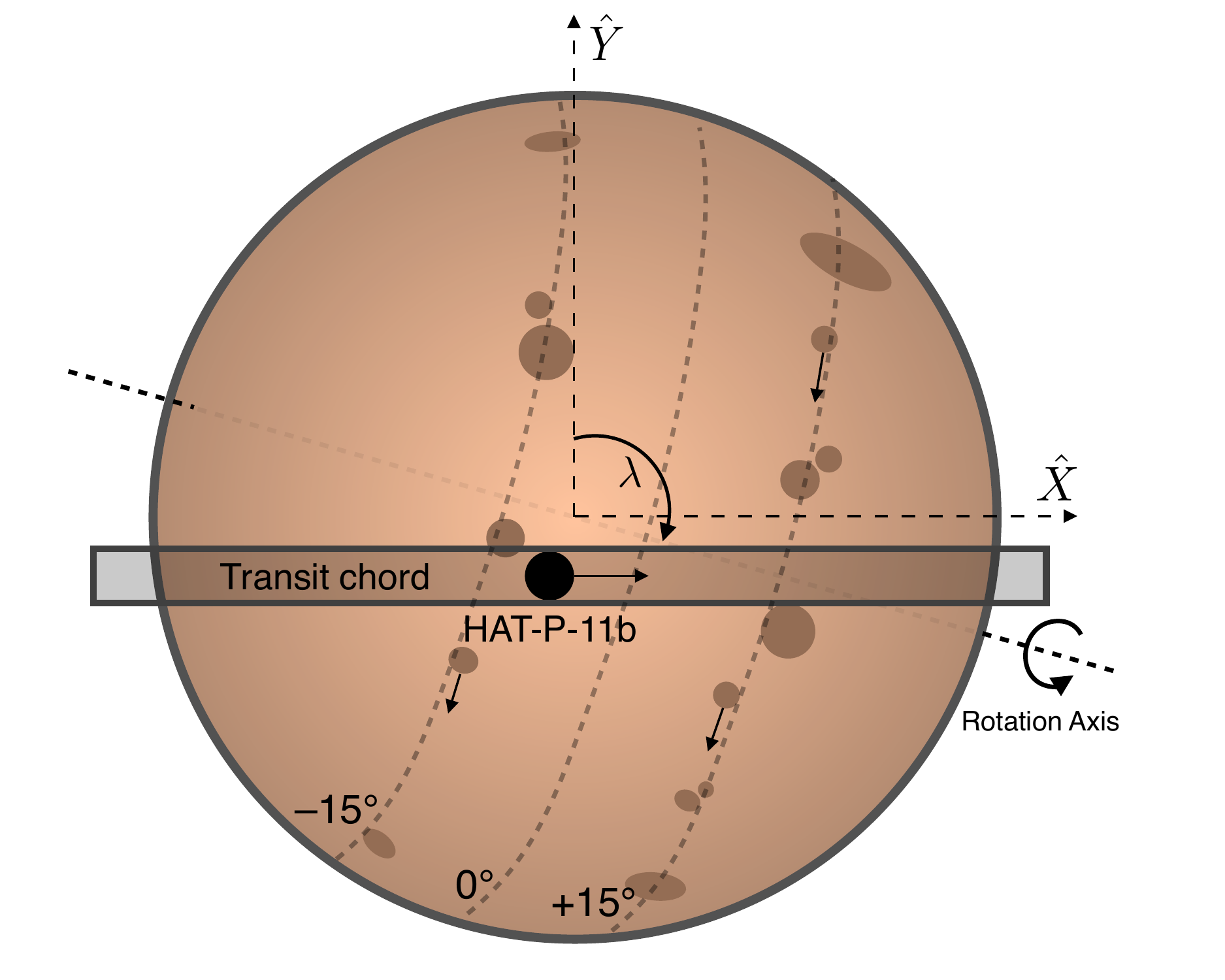}
\caption{The HAT-P-11 system in the observer-oriented coordinate system of \citet{Fabrycky2009}, roughly to scale. The planet's orbit is misaligned from the stellar rotation axis by the projected spin-orbit angle $\lambda = 106^\circ$ \citep{Sanchis-Ojeda2011}, and the north rotational pole of the star is inclined away from the observer by $i_s = 100^\circ$.}
\label{fig:schematic}
\end{figure}

We observe starspot occultations as positive flux anomalies during transit events. The amplitudes, durations and timing of the spot occultations constrain the spot locations and radii. If the starspot is a uniformly dark circular region on the star, and the planet passes over the edge of the spot in a grazing occultation, the resulting flux anomaly is an inverted ``v'' shape, analogous to the shape of an inverted eclipsing binary light curve. If the planet completely occults the spot or the spot completely circumscribes the planet, the resulting flux anomaly is an inverted ``u'' shape, like an inverted exoplanet transit event. There are many more grazing spot occultations (``v''-shaped, roughly approximated by Gaussians) than complete spot occultations. In most \kepler\ transits of HAT-P-11, there are between one and four spot occultations with amplitudes more than a few times the noise.

It is notoriously difficult to measure starspot positions robustly, because they are described by several degenerate quantities. For example, the occultation of a small, very dark spot is often degenerate with a larger spot of less extreme intensity contrast. These degeneracies can be broken for host stars of transiting exoplanets like HAT-P-11. The orientation of the planet's orbit is measured from the transit light curve, and the orbital phase of the planet at each time maps to a position on the projected stellar surface that is being occulted. We can measure the orientation of the star with two angles -- the spin-orbit angle which is measured via the Rossiter-McLaughlin effect, and the stellar inclination. We need to assume a stellar inclination, since the spectroscopic $v\,\sin{i}$ is consistent with zero \citep{bakos2010} due to this star's long rotation period. \citet{Sanchis-Ojeda2011} found that active latitudes are evident in the spot positions recovered from the \kepler\ photometry, and they measure the stellar inclination by assuming that the active latitudes are symmetric with respect to the stellar equator. Using the same technique, we can then map the flux measured at a given time to the brightness of the stellar surface at a particular latitude and longitude. Then when the planet occults a dark starspot, the timing and shape of the positive flux anomaly in the transit light curve can be transformed into the position and radius of the starspot. 

Figure~\ref{fig:schematic} depicts the orientation of the system. HAT-P-11 b's orbit normal is nearly perpendicular to the host star's spin -- in other words, it nearly orbits over the host star's poles. Thus each transit cuts a chord across the stellar surface from pole to pole, across most latitudes and over a narrow range in longitude. The transit chords start near the southern rotational pole of the star in the eastern hemisphere, pass over the sub-observer meridian in the northern hemisphere, and end to the northwest of where the chord began. The most complete latitude coverage is from the equator to $\pm 50^\circ$, with no transits occulting near the poles. The north rotational pole of the star is tilted into the sky-plane by $\sim10^\circ$.

We cannot say definitively whether or not the starspots of HAT-P-11 are occulted in consecutive transits. Since the stellar rotation period is roughly $P=29.2$ d and the orbital period is $P = 4.8878$ days, the planet occults the same longitude once per stellar rotation, which appears to be longer than the lifetime of spots on HAT-P-11 \citep{Sanchis-Ojeda2011} and similar to the lifetimes of sunspots \citep{Solanki2003}. The search for repeated spot occultations is made more difficult by the fact that there are active latitudes on the star, so one would expect to find spot occultations at similar orbital phases in each transit. We therefore assume that each spot occultation belongs to only one spot, and fit each transit light curve independently.

Starspot photometry models are difficult to optimize. The starspot occultations impart only small anomalies to a few flux measurements per transit, so the region of spot latitude-longitude-radius space that produces an improvement in likelihood is often very small and computationally expensive to find with a blind search. Preliminary experiments by \citet{Hebb2017} showed that unseeded MCMC fits required very long integration times to fully explore the parameter space before converging into likelihood maxima. However, the spot occultations of HAT-P-11 have quite high signal-to-noise owing to the star's brightness ($K_p = 9.17$), which makes them relatively simple to locate using peak-finding algorithms. We therefore devise a heuristic spot occultation model in Section~\ref{sec:friedrich}, which provides us with sensible initial conditions for the full \stsp\ forward model, which we discuss in Section~\ref{sec:stsp}.

\subsection{Initial, Heuristic Spot Occultation Model} \label{sec:friedrich}

We need initial guesses for spot positions in stellar latitude and longitude, and the stellar inclination angle $i_s$ to seed our \stsp\ model. We find spot occultations in the transits in a two-step process. First, we subtract the light curve by the transit model from Section~\ref{sec:transit}, which produces residuals near zero except near spot occultations. We then smooth the flux residuals by convolving them with a Gaussian kernel, and apply a local-maximum peak-finding algorithm to find the times and amplitudes of spot occultations in the residuals. We exclude any peaks detected within 5\% of the transit duration of ingress or egress, since we are not able to measure reliable spot properties for these highly foreshortened spots near the stellar limb. 

We approximate the residuals of each transit as the sum of Gaussian perturbations, with one Gaussian per spot. We marginalize over the Gaussian amplitude, mid-occultation time and width using the affine-invariant MCMC method. We assign a positive prior to the amplitude to search only for occultations of dark spots, and a flat prior to the mid-spot occultation time to exclude spots occulted within 5\% of the transit duration from ingress or egress. We apply a flat logarithmic prior to the spot-occultation width $\sigma$ to include only real occultations of small spots. We set priors on the spot occultation width to limit our spots to the regime $1.5 < \sigma < 8.6$ minutes -- the lower limit prevents the model from choosing very narrow Gaussians that affect single fluxes, which are typically outliers. The upper limit of the prior prevents the model from choosing very long duration spot occultations, which we do not observe in the \kepler\ data. We also apply a significance cut which excludes any spot occultations with significance $\Delta$BIC$ < 20$. This yielded 294 spots, on 138 of the 205 complete transits in the full \kepler\ light curve. 

We also ran a null test to verify that false-positives are not being incorrectly identified as spots. We offset the mid-transit time by one quarter of an orbital phase and set $R_p/R_\star = 0$ to search for false-positive spot-occultations in regions of the light curve where no transit is occurring. If there was significant correlated noise in the HAT-P-11 light curve with amplitudes and time-scales similar to the spot-occultation signals, those fluctuations would be detected as candidate spot occultations. No such false-positive spot occultations were detected by the peak-finding algorithm. We therefore conclude that correlated noise is not a significant source of false-positive detections of spot occultations on the scales relevant to this work.

\subsection{Stellar Inclination} \label{sec:i_s}

The starspot positions that we extract depend on the stellar orientation that we assume when computing the spot positions. Two angles define the orientation of the stellar rotation axis: (1) the stellar inclination $i_s$, which is the angle between the observer, the center of the star, and the rotation axis of the star; and (2) the projected spin-orbit angle $\lambda$, which is the tilt of the stellar rotation axis on the sky-plane with respect to the orbit normal of the planet. See Figure 1 of \citet{Fabrycky2009} for a graphical representation of these angles. 

The projected spin-orbit angle $\lambda$ has been constrained with the Rossiter-McLaughlin effect \citep{Winn2010, Hirano2011}, but the stellar inclination $i_s$ is more difficult to measure. In principle, it can be calculated for systems with known stellar rotation periods and spectroscopic rotational velocities ($v \sin i_s$), but \citet{bakos2010} found only a weak constraint on the projected rotational velocity. 
Using a different approach, \citet{Sanchis-Ojeda2011} noted that the distribution of starspots on HAT-P-11 resembled active latitudes like those of the Sun. The authors fitted the spot latitude distribution to solve for the stellar inclination by requiring the active latitudes to be symmetric across the stellar equator. They discussed two possible stellar orientations that explain the apparent active latitudes which they called the ``pole-on'' and ``equator-on'' solutions. In this paper, we reject the ``pole-on'' solution, because high-latitude spots viewed from a pole-on orientation would not move into and out of view sufficiently to produce the observed $\sim 3$\% rotational variability. We adopt the ``equator-on'' solution hereafter.

\citet{Sanchis-Ojeda2011} estimated the stellar inclination with observations from \kepler\ Quarters 0-2. Here, we carry out a similar analysis with the complete \kepler\ light curve from Quarters 0-17, which yields a stronger constraint on the stellar inclination. We procede by constructing a probabilistic model for the distribution of the spot latitudes. We model the probability distribution of spots as a function of latitude using a Gaussian mixture model $p(\ell)$, which is the sum of two normal distributions with mean latitudes $m_1$ and $m_2$, standard deviation $\sigma$, and relative amplitudes $a$ and $(1 - a)$,
\begin{multline}
p(\ell) = a \, \exp(-\frac{(\ell - m_1)^2}{2\sigma^2}) +\\ (1-a) \, \exp(-\frac{(\ell - m_2)^2}{2\sigma^2}).
\end{multline}
The time-dependent mean latitudes $m_1$ and $m_2$ are
\begin{eqnarray}
m_1(t) &=& \bar{\ell} + \ell^\prime t + \Delta i_s \\ 
m_2(t) &=& -(\bar{\ell} + \ell^\prime t) + \Delta i_s 
\end{eqnarray}
where the mean latitude is $\pm \bar{\ell}$,\footnote{We use the symbol $\ell$ to represent stellar latitudes, rather than $\lambda$ as is used in the sunspot literature, to avoid confusion with the projected spin-orbit angle, which by the convention of \citet{Ohta2005} is also called $\lambda$.} and $\Delta i_s$ is the difference between the stellar inclination measured by the probabilistic model and the stellar inclination published in \citet{Sanchis-Ojeda2011}. 

We allow the mean latitudes $m_1$ and $m_2$ to vary in time since the Sun's active latitudes migrate from high to low latitudes throughout the solar activity cycle. The parameter $\ell^\prime$ therefore tests whether or not we can detect evolution in the mean spot latitudes throughout the four years of the \kepler\ mission. The Sun's activity cycle is $\sim 11$ years long, and significant migration in mean spot latitude can be detected over four year intervals. If the activity cycle of HAT-P-11 is long compared to four years, the slower latitude evolution could be reflected in small values of $\ell^\prime$.

We force the mean latitudes to be symmetric about the stellar equator, and allow the northern and southern hemisphere distributions to have independent amplitudes. We assume the distribution is symmetric about the equator because: (1) on few-year time-scales the mean latitudes of the solar active latitudes are approximately symmetric; and (2) we have no more robust measurement of the stellar inclination to assert that the active latitudes are asymmetric. 

We maximize the likelihood of the observed distribution of spot latitudes from our simple model for values of $a, \sigma, \bar{\ell}, \ell^\prime$ and $\Delta i_S$ with the MCMC package \texttt{emcee} \citep{Foreman-Mackey2013}. We find the maximum-likelihood slope of the mean active latitudes is $\ell^\prime = 0.9 \pm 0.8$ degrees per year, consistent with no latitude evolution. This may indicate that the activity cycle of HAT-P-11 is long compared to four years. Since there is no evidence for  time-evolution of the active latitudes, we fix $\ell^\prime = 0$ and fit the model again.

The maximum-likelihood solution for the stellar inclination is $i_s = 100 \pm 2^\circ$, following the angle definition in \citet{Fabrycky2009} ($i_s$ is the angle between the observer's line of sight, the center of the star, and the stellar rotation axis), which we adopt as fixed throughout the rest of this work. This stellar inclination angle is consistent with the inclination reported in \citet{Sanchis-Ojeda2011}, though the values differ due to their choice of coordinate system. We will revisit the distribution of spot latitudes with solutions from the more detailed spot model in Section~\ref{sec:lat_dist}.

\subsection{Spin-orbit misalignment} \label{sec:obliquity}

We can measure the obliquity -- or the de-projected spin-orbit misalignment -- of HAT-P-11 with our revised measurements of $i_o$ and $i_s$. We solve Eqn.~9 of \citet{Fabrycky2009} for the obliquity $\psi$
\begin{equation}
\cos \psi = \sin i_s \cos \lambda \sin i_o + \cos i_s \cos i_o,
\end{equation}
and find $\psi \approx 106^\circ$, consistent with the obliquity reported by \citet{Sanchis-Ojeda2011}. This provides another check on our coordinate system which follows the definitions of \citet{Fabrycky2009} and differs from \citet{Sanchis-Ojeda2011}, but yields the same obliquity angle. 

\subsection{Spot contrasts} \label{sec:contrast}

\begin{figure}
\centering
\includegraphics[scale=0.35]{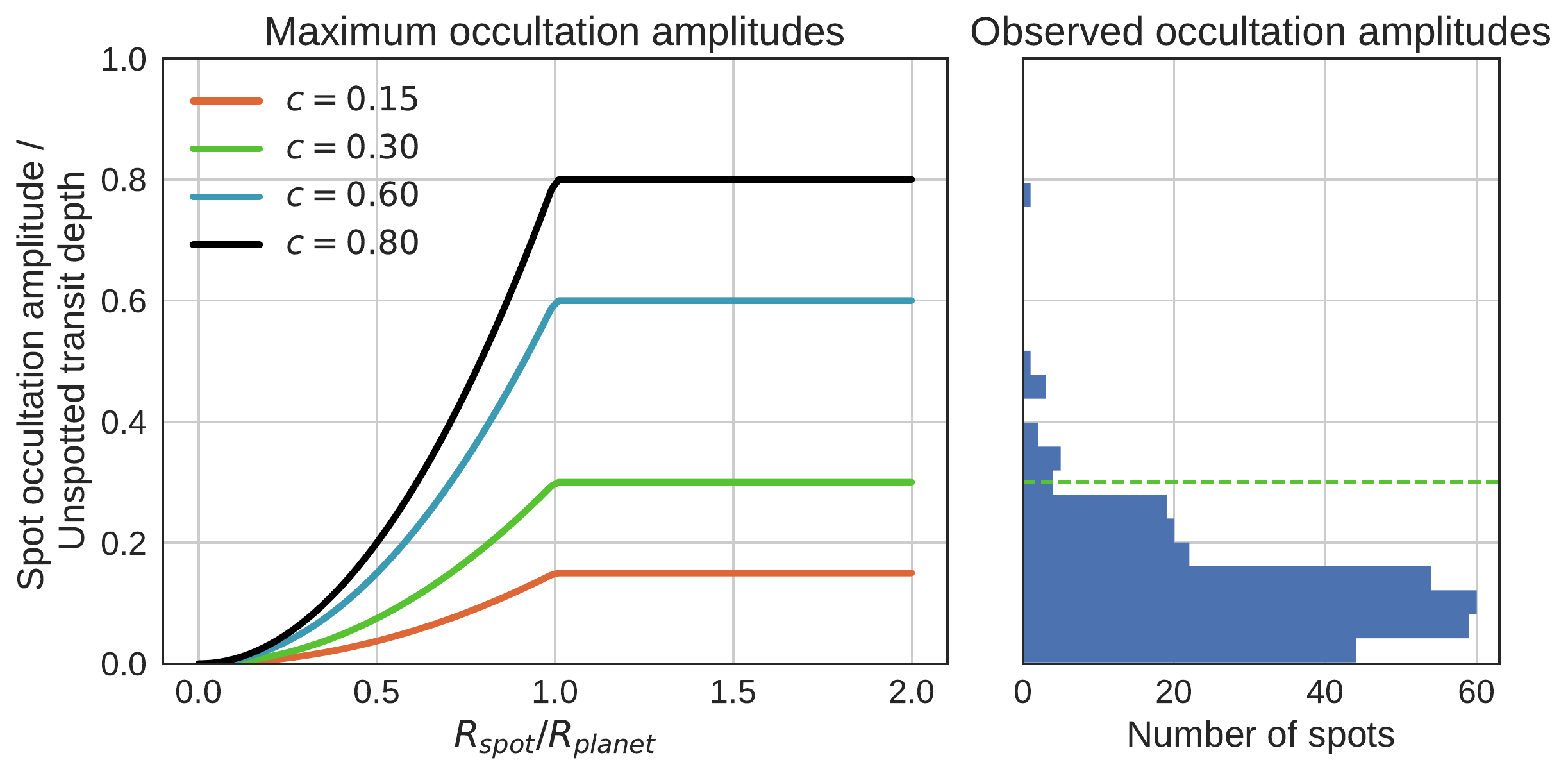}
\caption{\textit{Left:} amplitudes of positive flux anomalies during spot occultations, normalized to the unspotted transit depth, as a function of the spot size and contrast. \textit{Right:} observed spot occultation amplitudes of HAT-P-11. 95\% of the spot occultations have normalized amplitudes $\le 0.3$ below the green dashed line, as one would expect from spots with the mean solar contrast $c=0.3$. The largest observed spot-occultation amplitude requires a spot contrast $c \ge 0.8$ -- similar to the spot contrast of sunspot umbra. Since most other spots are consistent with the mean solar spot contrast $c = 0.3$, we adopt the solar contrast value for the spots of HAT-P-11.}
\label{fig:contrast}
\end{figure}

In this work, \stsp\ approximates starspots as circular features with homogeneous contrast. We can define the spot intensity contrast relative to the local photosphere $c$ as
\begin{equation}
c = 1 - I_{spot}/I_{phot} \label{eqn:contrast}
\end{equation}
where $I_{spot}$ is the mean intensity inside the dark spot, $I_{phot}$ is the intensity of the local photosphere, and $I_{spot} < I_{phot}$. Spots with temperatures and intensities similar to the local photosphere $I_{spot} \approx I_{phot}$ are ``low contrast'', i.e. $c \rightarrow 0$, and spots with extreme temperature differences $I_{spot} \ll I_{phot}$ are ``high contrast'' and $c\rightarrow 1$. High-resolution studies of sunspots show that the spot darkness correlates with magnetic field strength in the vertical component \citep{Keppens1996, Leonard2008}.

Sunspots have complicated substructures each with their own contrast, such as the dark umbra and less dark penumbra. We cannot typically resolve such substructure in occultation photometry, so we chose to adopt the area-weighted contrast of the penumbra and umbra as the contrast for the entire spot. We can approximate sunspots as homogeneous circular features if we average over the penumbra and umbra, which have contrasts $c_{umbra} \sim 0.5 - 0.8$ and $c_{penumbra} \sim 0.15-0.25$. The mean area encompassed by the penumbra is roughly four times larger than the umbral area \citep{Solanki2003}. Adopting $c_{umbra} = 0.65$ and $c_{penumbra} = 0.2$, the area-weighted mean spot contrast of sunspots is $c_{total} = 0.3$. 

We can compare solar spot contrasts to constraints on the spot contrasts of HAT-P-11 from \kepler\ photometry. Spot contrasts are constrained by the amplitudes of spot occultations events. The difference in flux during a transit with a spot occultation and a transit without a spot occultation is set by the spot contrast, and the projected size of the spot compared to the planet. We derive spot occultation amplitudes as a function of spot radius and contrast in Appendix~\ref{sec:appendix_contrast}.

In Figure~\ref{fig:contrast} we compare the spot occultation amplitudes normalized by the flux of the unspotted star at each time during the transit, for a variety of spot contrasts and spot sizes with the observed spot amplitudes. As the spot contrast $c$ increases and the spot becomes darker, the amplitude of the spot occultation increases for spots of any radius. For spots larger than the planet, the contrast controls the maximum occultation amplitude. Therefore the maximum observed spot occultation amplitude sets a lower bound for the maximum spot contrast. The spot occultation with the largest normalized amplitude $\sim 0.80$ requires a spot contrast of $c_{min} \ge 0.8$, which is similar to the contrast of sunspot umbra. 95\% of the spot amplitudes could be produced by occultations of spots with the area-weighted mean solar spot contrast $c = 0.3$, so we adopt $c = 0.3$ as our spot contrast in fits with \stsp\ model, since it is consistent with both the spots of the Sun and HAT-P-11.

\section{Detailed \stsp\ Spot Occultation Model} \label{sec:stsp}

The \stsp\ model is constructed as follows \citep[see][for more details]{Hebb2017}. The star is represented by a series of discrete concentric circles with intensities decreasing radially outward to approximate limb darkening. Spots on the star are represented as non-overlapping circles that are darker than the local photosphere, which follow the stellar surface in fixed-body rotation. Each spot is defined by four parameters: radius, latitude, longitude, and intensity contrast relative to the photosphere. The planet is represented by an opaque circle, and the relative flux received by the observer is calculated throughout the orbit of the planet. We marginalize over the spot position and radius parameters.

We fit for spot properties with \stsp\ using the number of spots and initial positions given by the simple model in Section~\ref{sec:friedrich}, which narrows the sample to 138 transits with highly significant spot occultations. We approximate stellar limb darkening with 40 concentric circles. We fix the spot contrast to $c=0.3$, which we justified in Section~\ref{sec:contrast}.

We run the affine-invariant MCMC with 300 chains and no priors for each transit, until the parameter posterior distributions are stationary \citep{Goodman2010}. One advantage of the pure \texttt{C} implementation of \stsp\ is that it is naturally portable and scalable for distributed computing. We run \stsp\ for each of the 138 transits independently on the Extreme Science and Engineering Discovery Environment (XSEDE) Open Science Grid \citep{osg, xsede}.

\subsection{Model parameter degeneracies}  \label{sec:degen}

\begin{figure}
\centering
\includegraphics[scale=0.2]{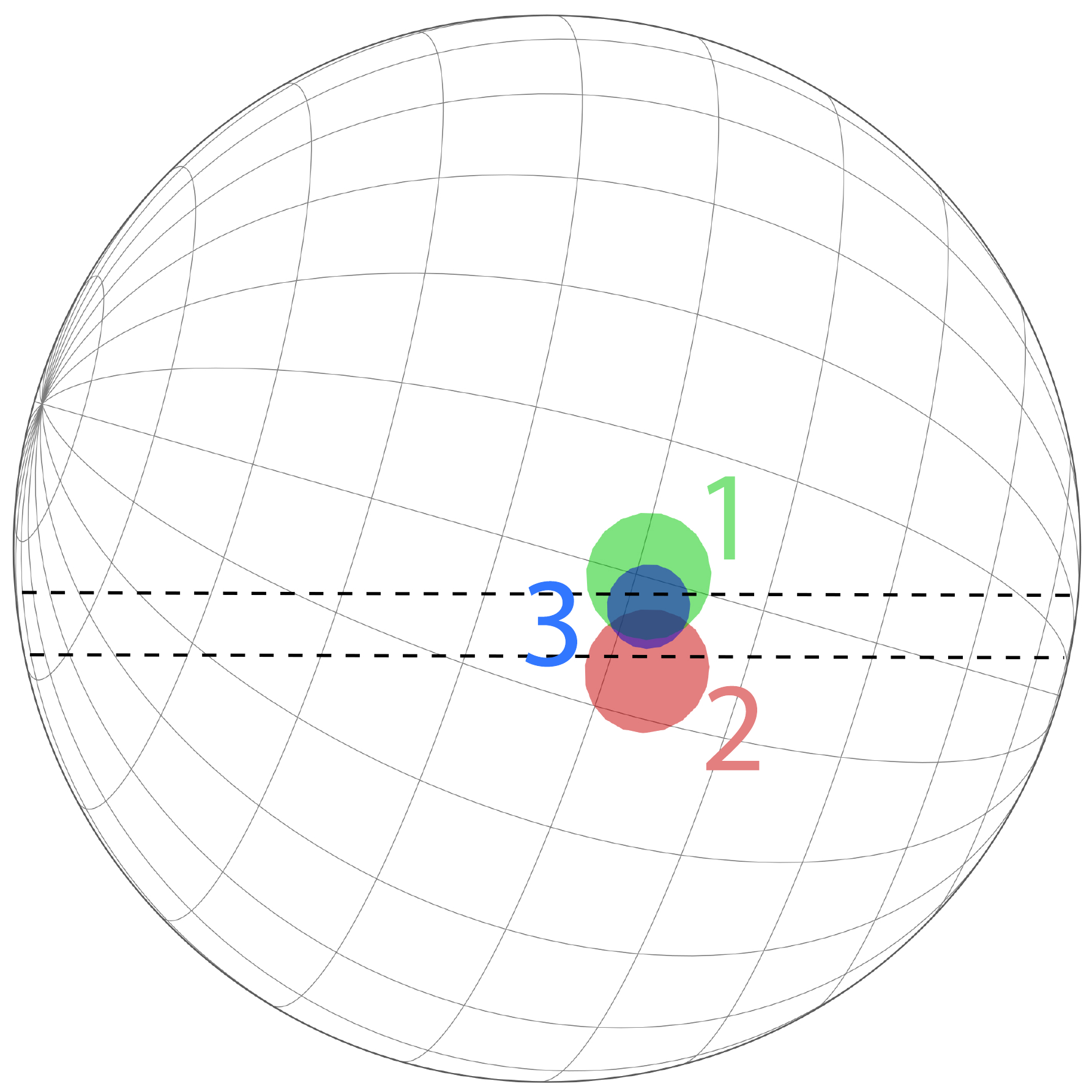}
\includegraphics[scale=0.45]{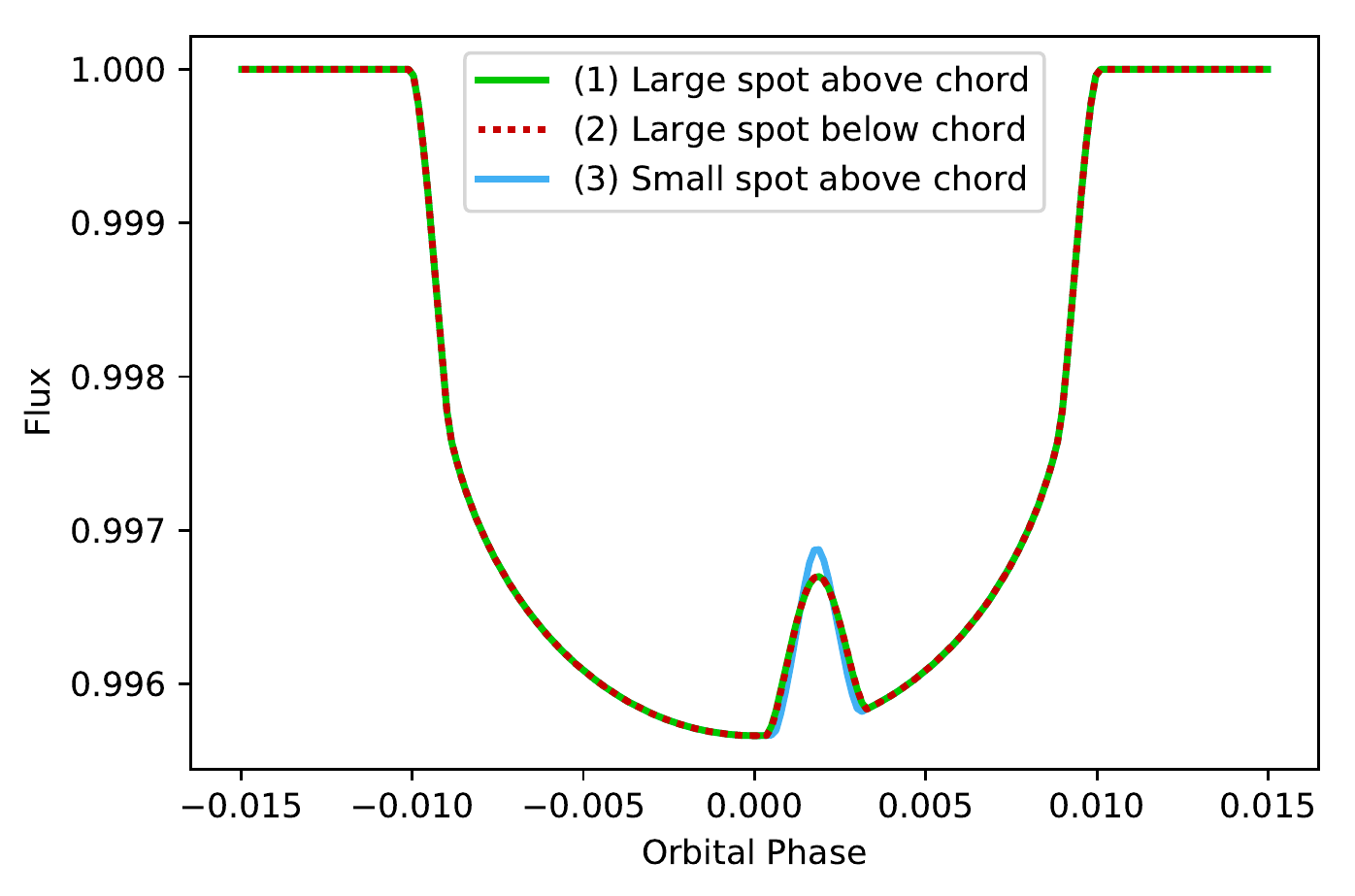}
\caption{\textit{Upper}: Map of a few hypothetical spots on HAT-P-11 which would produce similar anomalies in the transit light curve.  The transit chord is bounded by the black dashed lines, the red latitudinal grid mark is the stellar equator -- the stellar rotational pole is tilted into the page and on the right. In this diagram, the planet transits from left to right, from near opposite the rotational pole to near the rotational pole. The ($R_{spot}/R_{star}$, latitude, longitude)  parameters for spot 1 (green), 2 (red), and 3 (blue) are: (0.12, 1.4$^\circ$, 359.8$^\circ$), (0.12, 4.3$^\circ$, 9.8$^\circ$), and (0.08, 2.3$^\circ$, 2.9$^\circ$), respectively. \textit{Lower:} \stsp\ model transits for each spot in the map above. With \kepler's flux precision for HAT-P-11 ($\sim 80$ ppm), these three models would be indistinguishable.}
\label{fig:rp_degeneracy}
\end{figure}

For most transit light curves with spot occultations, there exists a series of spot positions and radii which produce equally good fits to the observations. There are two main degeneracies in our choice of spot model which are critical to understanding the fit results of HAT-P-11; we will call these degeneracies: (1) the transit chord degeneracy and (2) the radius-position degeneracy. 

The transit chord degeneracy is a simple consequence of symmetry. For any small spot placed near the transit chord, a spot of the same radius could be placed on the opposite side of the transit chord (at the same distance from the transit chord) to create an identical bump in the light curve. See for example spots 1 and 2 in Figure~\ref{fig:rp_degeneracy}. 

The transit chord degeneracy may be broken in two scenarios: (1) for some star-planet systems with large impact parameters, the spots would be significantly more foreshortened on one side of the transit chord than the other; or (2) large spots that subtend large angles from the center of the stellar disk to the limb will be more foreshortened near the limb than at disk center, producing asymmetries between spot-crossing ingress and egress. HAT-P-11 has impact parameter $b = 0.141$ so spots projected onto either side of the transit chord will appear roughly symmetric, and therefore the spot position solutions most often come in pairs that are symmetric about the transit chord. However, there are a few exceptionally large spots that give rise to asymmetric spot crossings, which allows the model to select a spot position on only one side of the transit chord.

The radius-position degeneracy arises from trade-off in spot occultation amplitude between spot size and position. A large spot which grazes the edge of the transit chord will produce a bump in the transit light curve similar to a much smaller spot laying within the transit chord. See for example spots 1 and 3 in Figure~\ref{fig:rp_degeneracy}. 

The radius-position degeneracy can be broken with observations at infinite time resolution and flux precision. In the \kepler\ observations of HAT-P-11, the one minute cadence and the single measurement uncertainty $\sigma_{\Delta F / F} \sim 80$ ppm prevent us from distinguishing between small spots near to the transit chord and somewhat larger spots farther from the transit chord. 

More details about \stsp\ model parameter degeneracies are discussed in \citet{Hebb2017}. Examples of \stsp\ fits to the HAT-P-11 light curves and the effects of these degeneracies are discussed in detail in the following section.

\subsection{Examples of degeneracies in results}

\begin{figure}
\centering
\includegraphics[scale=0.2]{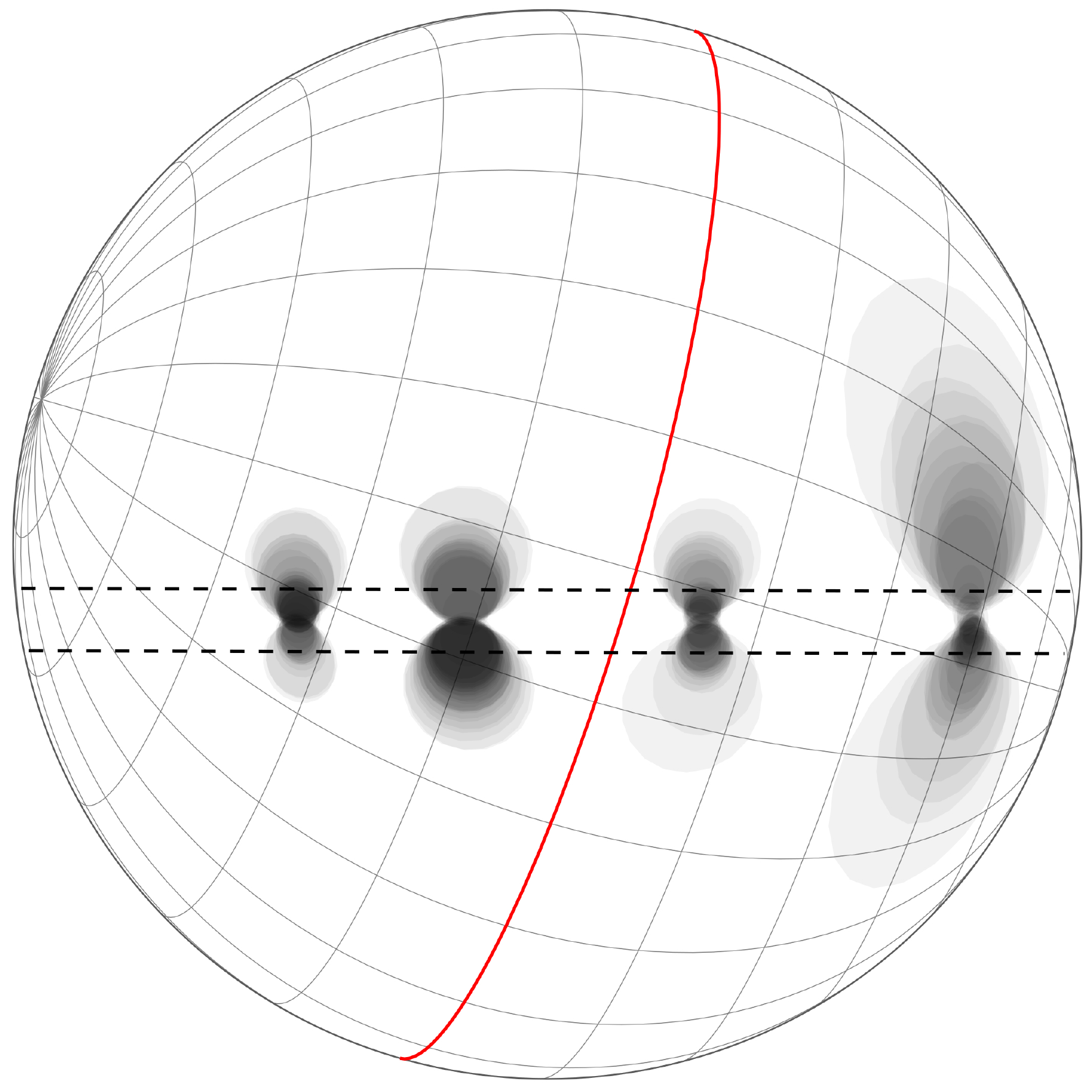}
\includegraphics[scale=0.45]{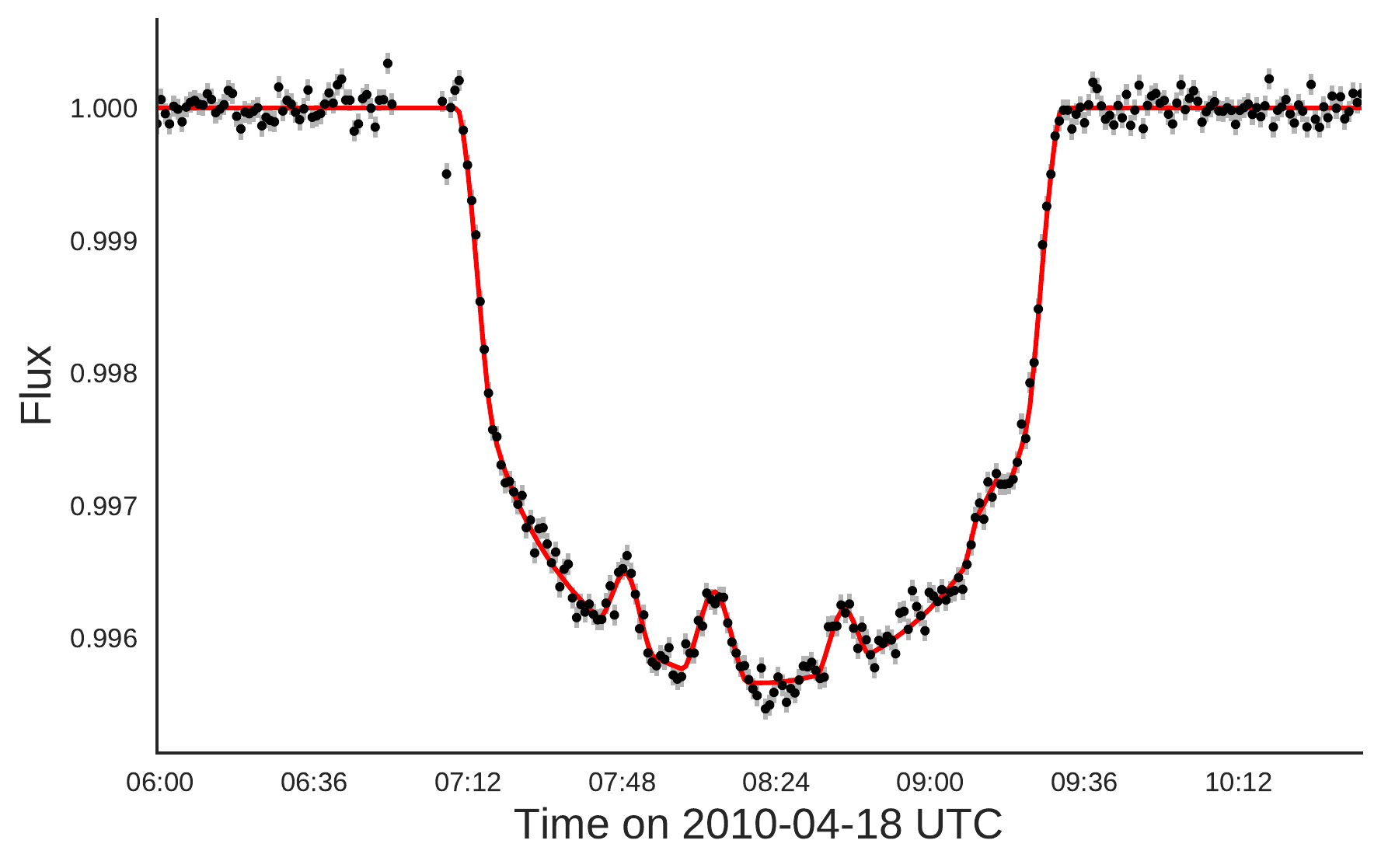}
\caption{\textit{Lower:} an example transit light curve of HAT-P-11 b (black points) with the maximum likelihood \stsp\ model (red curve). \textit{Upper}: a few random draws from the posterior samples for the spot positions and radii. These four spots are highlighted in blue on the spot map in Figure~\ref{fig:map}.}
\label{fig:transit_063}
\end{figure}

\begin{figure}
\centering
\includegraphics[scale=0.45]{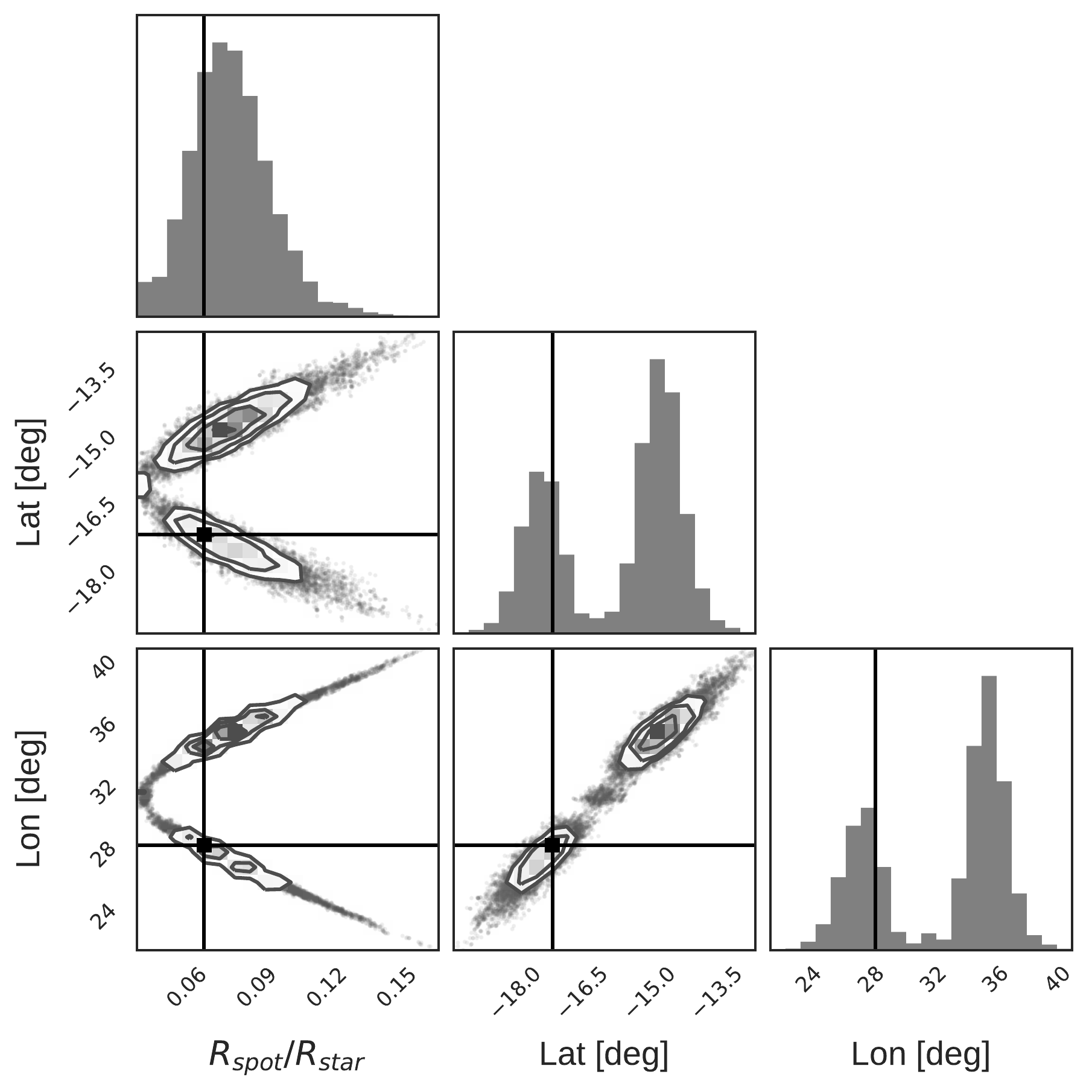}
\caption{Correlations between posterior samples for latitude, longitude and spot radius for the spot second from the left in the graphic in Figure~\ref{fig:transit_063}. The vertical/horizontal black lines mark the maximum likelihood values from the Markov chains. For the analysis of the spot latitude and radius distributions, we use the maximum likelihood values to compute the spotted area in the transit chord, for example, which selects one of the two possible groups of solutions for each spot. Note that the model constrains the minimum spot radius for this spot despite the position-radius degeneracy, which is a result of fixing the spot contrast.}
\label{fig:corner_063}
\end{figure}

We can begin to understand the spot radius-position degeneracy which affects the radius distribution by inspecting the transit on April 18, 2010; see Figures~\ref{fig:transit_063}  and \ref{fig:corner_063}. There are four spot occultations visible in the transit light curve, so we seed the \stsp\ model with four spots, and optimize for the latitude, longitude and radius of each spot with fixed flux contrast ($c=0.3$, as defined in Equation~\ref{eqn:contrast}). The posterior samples for latitude, longitude and radius of each spot cluster into two groups of solutions -- one on each side of the transit chord. In some spot occultations, asymmetry in the photometry produces a preferred solution on one side of the transit chord. In the case of the spot posteriors shown in Figure~\ref{fig:corner_063} (see also the light curve and spot geometry in Figure~\ref{fig:transit_063}), the latitude and longitude have bimodal solutions. Therefore, rather than adopting the mean of these bimodal posteriors as the best solution, we use the parameter values at the maximum likelihood step of the MCMC chains to study spot radii (and latitudes).

For a fixed spot contrast, the radius-position degeneracy biases us towards larger radii. The asymmetry towards large radii can be seen in the bottom left plot of Figure~\ref{fig:corner_063}. If the spot contrast cannot vary, there exists a minimum spot radius which is capable of reproducing the observed occultation amplitude for a direct spot occultation (impact parameter $b=0$). Any indirect or grazing occultations ($b \neq 0$) would require a larger spot to produce the same occultation amplitude, producing an abundance of possible solutions with large spots, centered farther away from the transit chord. For this reason, we do not assert that any spots on HAT-P-11 are certainly larger than the largest sunspot, though the maximum likelihood solutions suggest such spots exist (more on spot radii in Section~\ref{sec:radii}).

\begin{figure}
\centering
\includegraphics[scale=0.2]{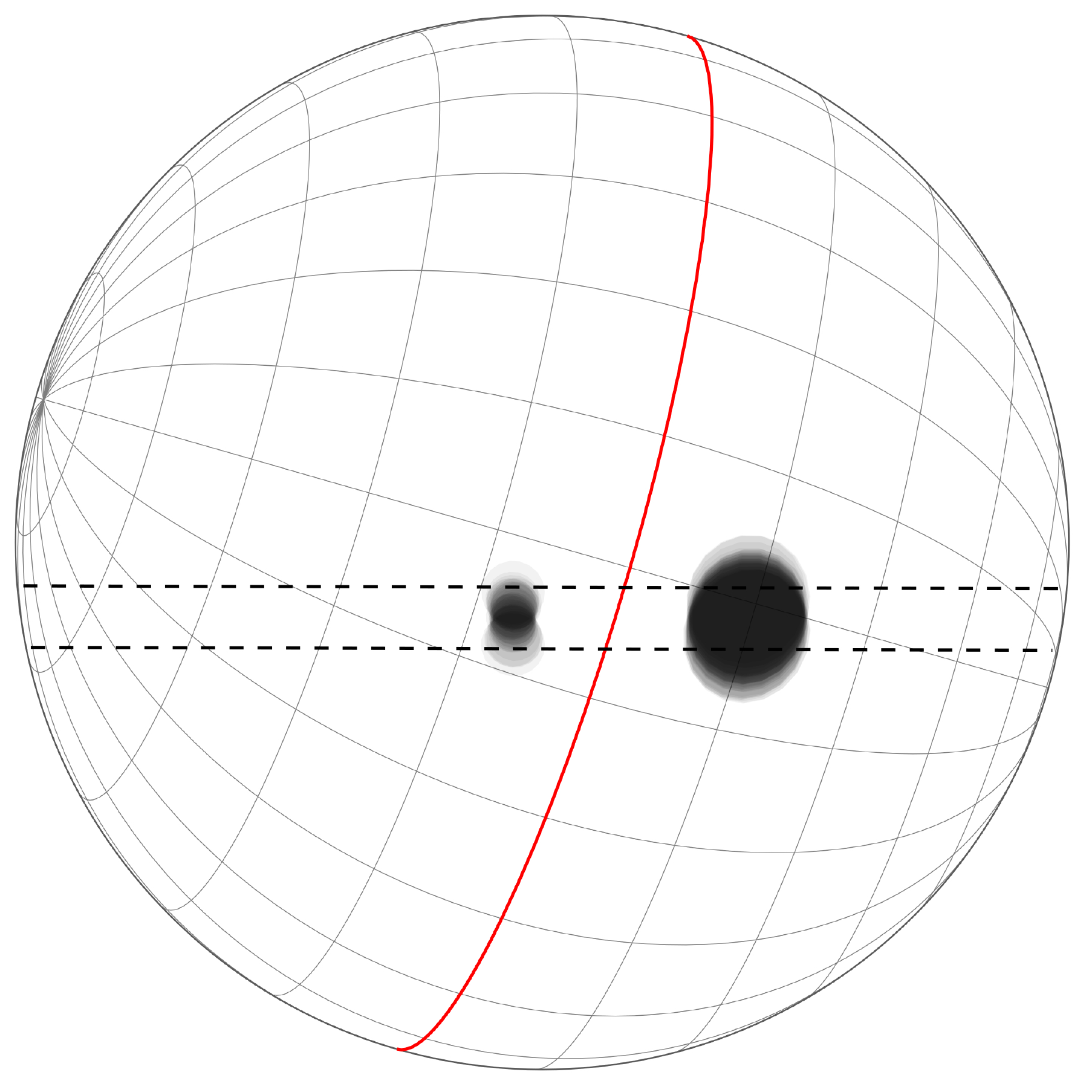}
\includegraphics[scale=0.4]{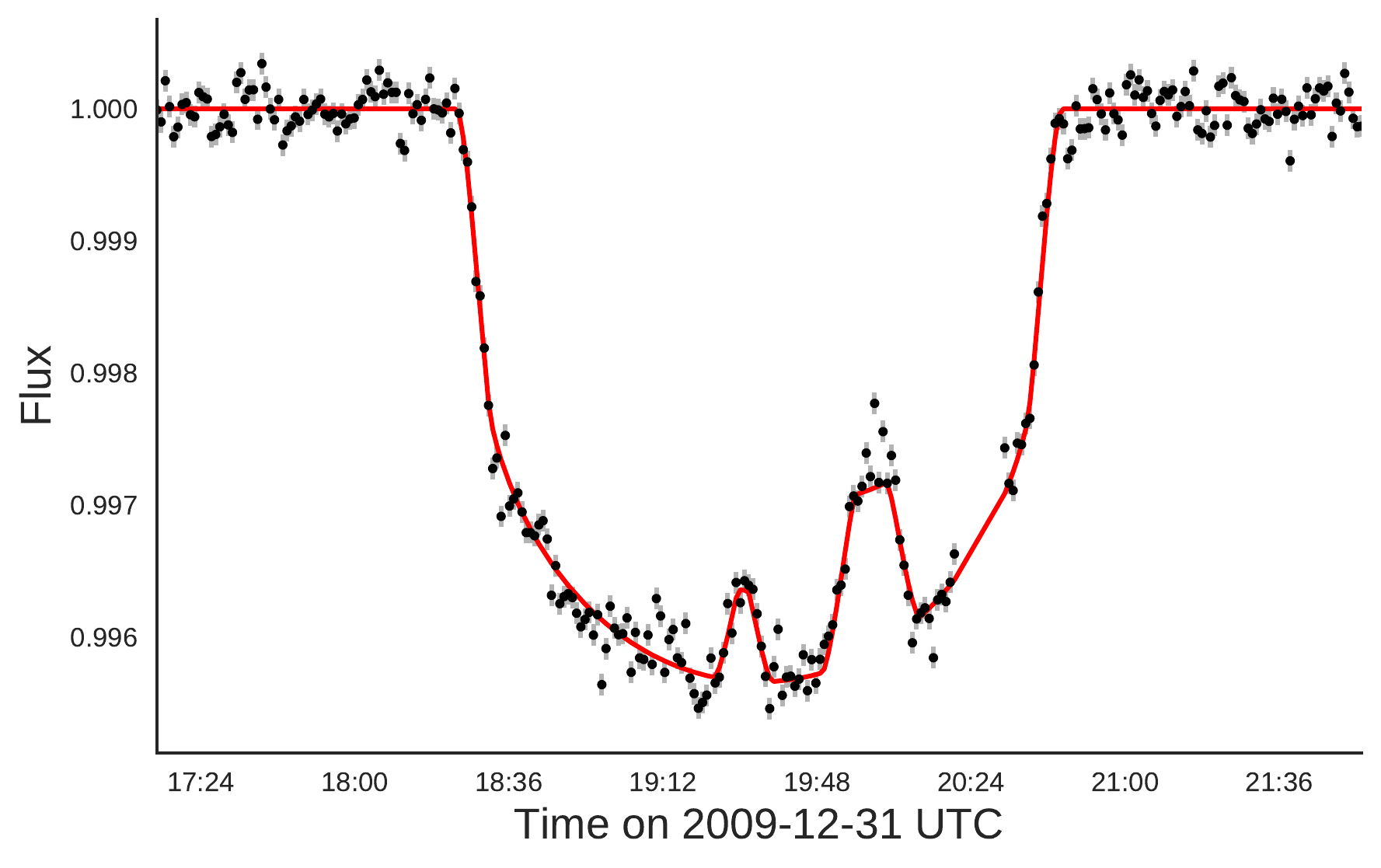}
\caption{\textit{Lower:} an example transit light curve of HAT-P-11 b (black points) with the maximum likelihood \stsp\ model (red curve). \textit{Upper}: a few random draws from the posterior samples for the spot positions and radii. The transit chord is bounded by the black dashed lines, the red latitudinal grid mark is the stellar equator -- the stellar rotational pole is tilted into the page and on the right. In this diagram, the planet transits from left to right, from near opposite the rotational pole to near the rotational pole. The grid lines are separated by 15$^\circ$, so the active latitudes appear near the grid lines on either side of the equator. Note that the flat-topped spot occultation model light curve corresponds to a spot that must appear to be wider than the transit chord -- setting a constraint on the minimum radius of the spot. The observed fluxes are higher than the model near the mid-occultation time, which is suggestive of a higher spot contrast near the center of the spot. These two spots are highlighted in red on the spot map in Figure~\ref{fig:map}. In physical units, these spots have radii of $22400^{+4000}_{-300}$ and $63700^{+3000}_{-200}$ km.}
\label{fig:transit_044}
\end{figure}

Figure~\ref{fig:transit_044} shows an example light curve model and a few draws from the spot parameter posteriors for the transit on December 31, 2009 UTC. The model of the second spot occultation in this transit has a flat top, and an amplitude similar to $\Delta F/ \delta \sim c = 0.3$, which implies that the spot was occulted with a small impact parameter, and that the spot radius was larger than the planet radius. The geometry of this eclipse parallels planetary transits with negligible limb-darkening, which produce a ``u''-shaped eclipse with a flat bottom. We note that the \kepler\ fluxes at the peak of the second spot occultation have a net positive scatter. This could imply that a more extreme spot contrast is justified at the center of the spot ($c<0.3$), where one might expect the umbra to be. The duration of the flat-topped spot occultation is proportional to the diameter of the spot, so the position and size of this spot are relatively well-constrained by the photometry. This is reflected by the uniformity of the posterior samples of the second spot in Figure~\ref{fig:transit_044} compared to the earlier grazing spot occultation, which is less constrained. The inverted ``v''-shape of the first spot occultation implies that the planet either grazed the spot at high impact parameter -- similar to planetary transits or binary eclipses with high impact parameters, which produce ``v''-shaped eclipses -- or that the spot is similar in size to the size of the planet. As you can see in the samples from the posteriors on the map in Figure~ \ref{fig:transit_044}, the model tends towards a spot centered in the transit chord, slightly smaller than the planet.

\section{\stsp\ Results} \label{sec:results}
\subsection{Spot Map of HAT-P-11}

\begin{figure*}
\centering
\includegraphics[scale=0.58]{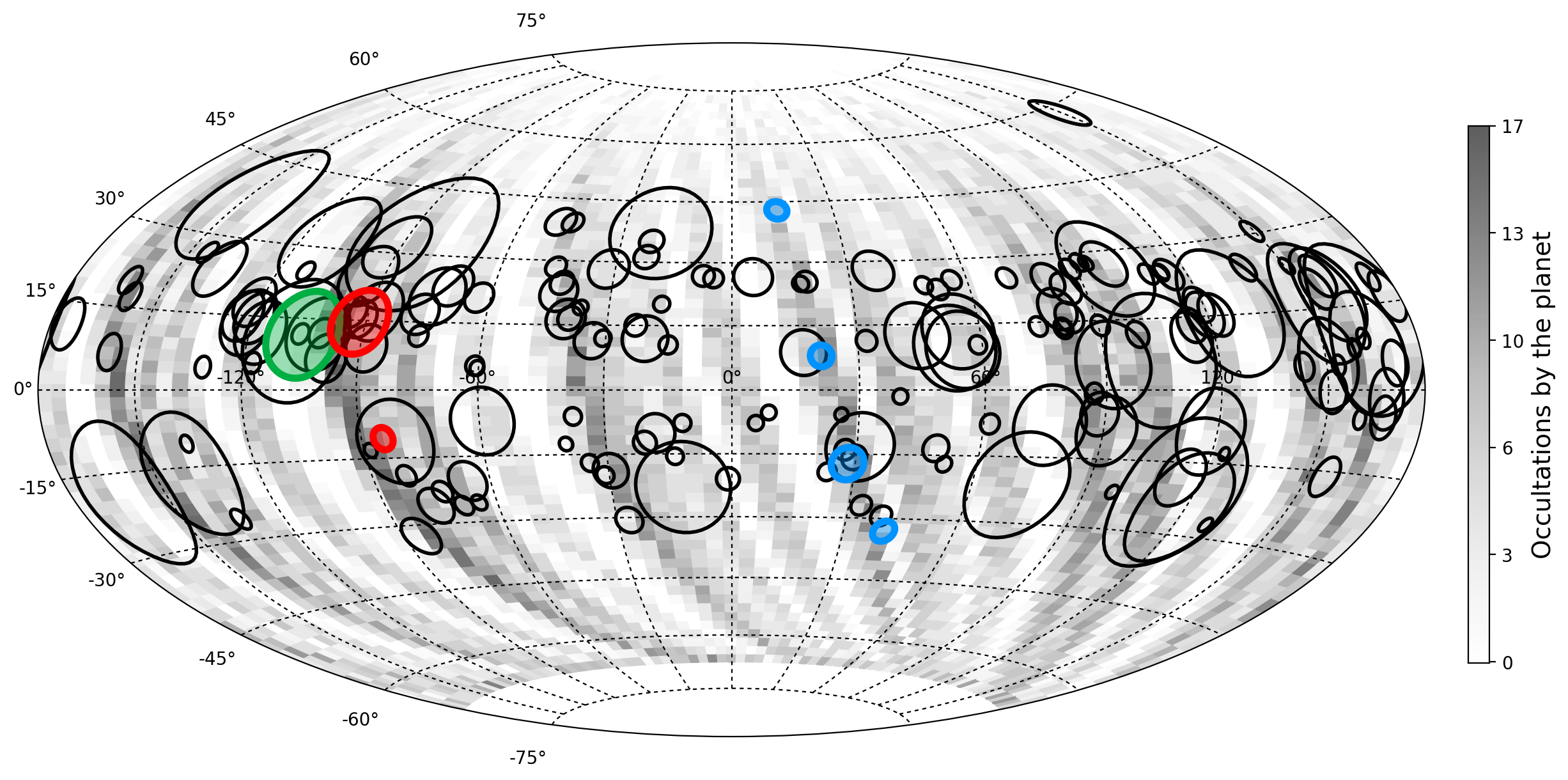}
\caption{Spots detected on HAT-P-11 with \stsp\ (see Section~\ref{sec:stsp}). The radius of each circle corresponds to the size of the spot. The shading beneath corresponds to the number of times the planet occulted that spatial bin on the stellar surface, which can be used as a proxy for relative completeness. Note that the spots occur preferentially at two active latitudes near $\pm15^\circ$. The 6:1 period commensurability between the orbital period and stellar rotation period produces the alternating longitudinal stripes in relative occultation number. The two red circles in the western hemisphere near longitude $-90^\circ$ highlight the spots derived from the transit light curve in Figure~\ref{fig:transit_044}, and the four blue circles in the eastern hemisphere near longitude $30^\circ$ correspond to the spots derived from the transit light curve in Figure~\ref{fig:transit_063}. The green circle near longitude $-100^\circ$ corresponds to the large spot discussed in Figure~\ref{fig:transit_071}.}
\label{fig:map}
\end{figure*}

We map the maximum-likelihood starspot positions for all 138 transits in Figure~\ref{fig:map}. The circles represent the positions and sizes of the spots inferred with \stsp. The shading of the map corresponds to the number of times the center of the planet occulted each location on the star, which is a proxy for completeness of the spot map -- darker regions were occulted more often. We choose to plot the maximum-likelihood spot positions and radii rather than the means of the posterior samples, because degeneracies between the model parameters can produce bimodal posterior distributions (see Section~\ref{sec:stsp} for discussion on model degeneracies).

The spin-orbit misalignment and spin-orbit commensurability of this system lead to highly inhomogeneous sampling in longitude, so an investigation into the true spot longitude distribution is beyond the scope of this work. However, asymmetries in spot latitude are detectible and readily visible in the spot map in Figure~\ref{fig:map}. The spots are distributed into two active latitudes near $\pm 16^\circ$ latitude, and the northern hemisphere appears to have more spots than the southern hemisphere. We investigate the latitude distribution of spots in the next section.

The transit chord of HAT-P-11 b is inclined $16^\circ$ from perpendicular to the stellar equator -- refer back to Figure~\ref{fig:schematic} for a schematic of the orientation. As a result of this slight misalignment from perpendicular, the planet never occults either pole of the star. It is possible that there are spots at latitudes $\ge 60^\circ$, which have been produced in simulations of highly-active sun-like stars \citep[e.g.][]{Schrijver2001}. Our spot map from transit photometry is insensitive to polar or high latitude spots. 

\subsection{Latitude Distribution} \label{sec:lat_dist}

\begin{figure}
\centering
\includegraphics[scale=0.48]{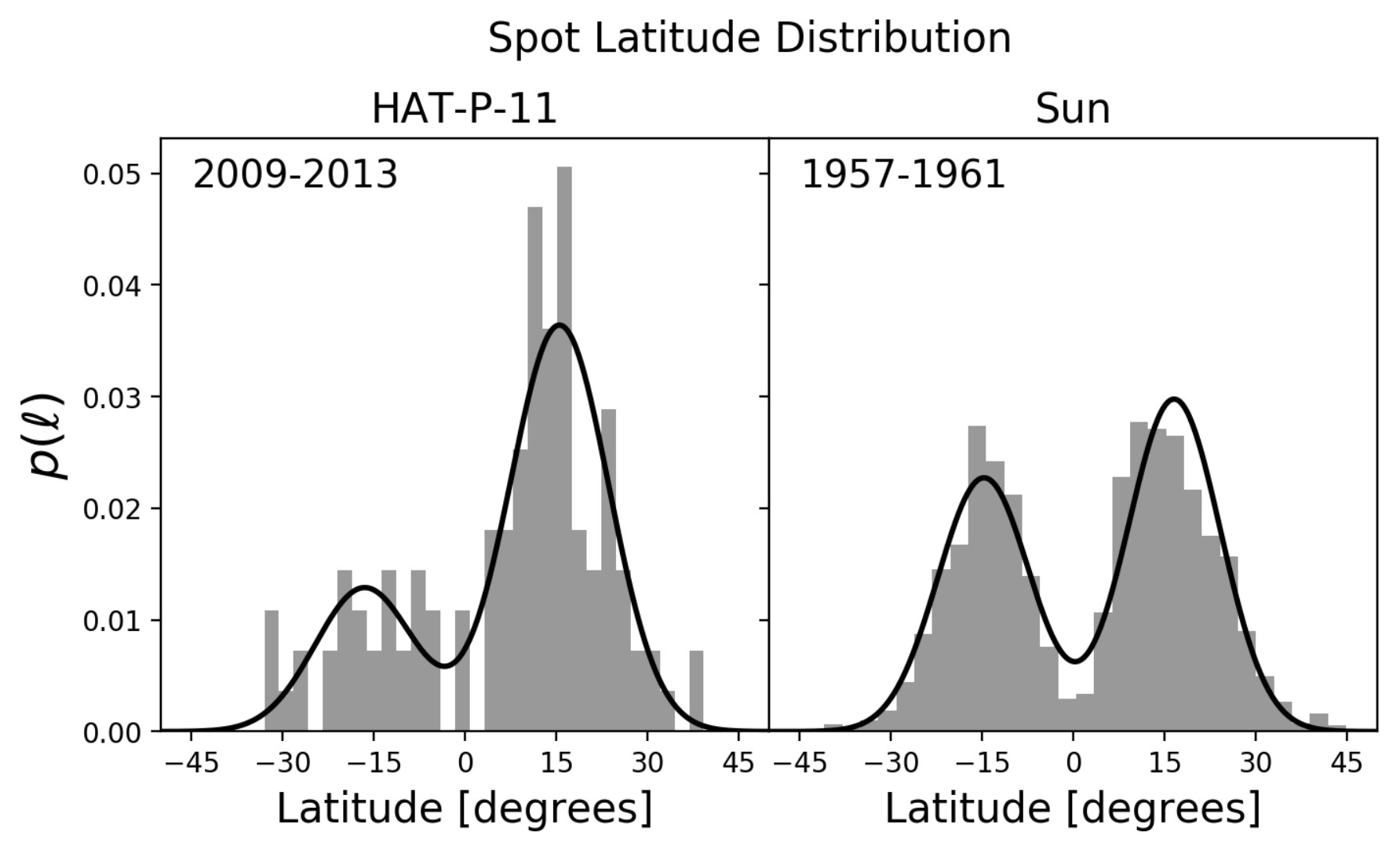}
\caption{Distribution of spot latitudes over four years of observations for both HAT-P-11 and the Sun. The four years of solar observations correspond to the maximum of solar Cycle 19 as observed by \citet{Howard1984}. The HAT-P-11 spot latitudes and their uncertainties are taken from the best-fit solutions from the \stsp\ spot occultation model. Both stars have active latitudes centered on $\pm 16^\circ$ with standard deviations of $\sim 8^\circ$. We put these latitude distributions in context throughout the solar activity cycle in Figure~\ref{fig:sun_vs_hat11}. Though HAT-P-11's hemispheric spot number asymmetry is greater than the Sun's in this particular bin of solar observations, we find that the asymmetry on HAT-P-11 is within the range observed on the Sun; see Section~\ref{sec:lat_dist} for details.}
\label{fig:latitude_model}
\end{figure}

\begin{figure}
\centering
\includegraphics[scale=0.6]{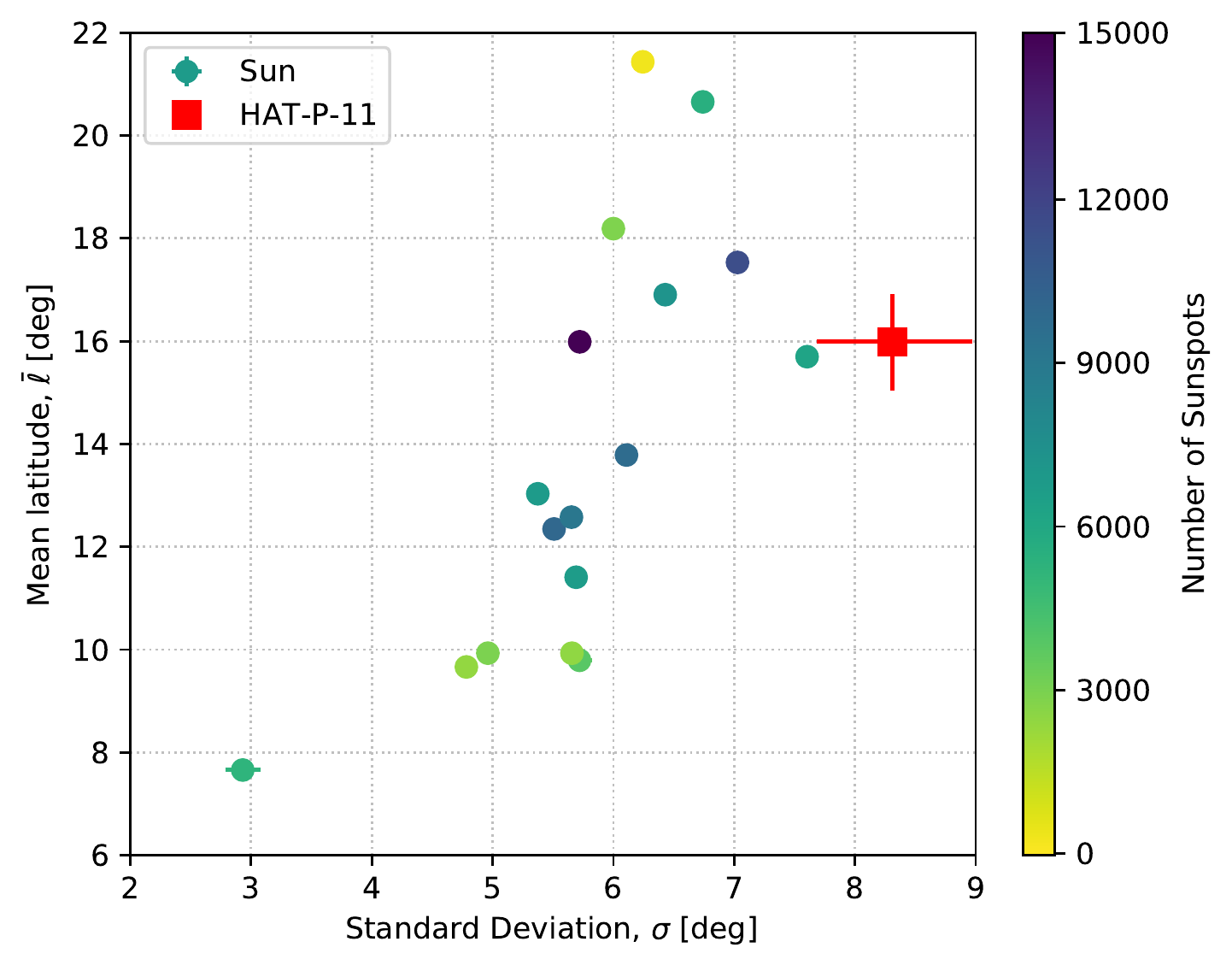}
\caption{Latitude distributions of sunspots and spots on HAT-P-11, parameterized by the mean latitude of spots in each hemisphere and the standard deviation of spot latitudes in each hemisphere. The circles are the best-fit parameters for four-year bins of the Mt.~Wilson sunspot catalog \citep{Howard1984}. The squares are fits to the HAT-P-11 latitude distribution over the four years of \kepler\ data. The colors of the solar circles represent the number of sunspots in each bin, which is a good proxy for phase of the solar activity cycle (darker points are nearer to solar maximum). Some of the fits have large uncertainties even with many spots, because the latitude distributions are not always well-approximated by Gaussians. The solar measurement closest to HAT-P-11's corresponds to the period 1957-1961 during solar Cycle 19, which is plotted in Figure~\ref{fig:latitude_model} for comparison to HAT-P-11.}
\label{fig:sun_vs_hat11}
\end{figure}

The mean latitudes of sunspots and the widths of their distributions across each hemisphere undergo an $\sim 11$ year cycle, which gives rise to the ``butterfly diagram'' of latitudinal spot density as a function of time \citep[see for example][]{Hathaway2011, Hathaway2015}. Near solar minimum, there are very few sunspots. Spots begin to appear at ``high'' latitudes $\left | \ell \right |  \sim 25^\circ$, and the mean spot latitude drifts towards the equator throughout the cycle, with the maximum number of spots occurring near $\left | \ell \right | \sim 15^\circ$. The northern and southern hemispheres of the Sun can have asymmetric numbers of spots, flares, and other activity indicators \citep[see for example][]{Newton1955, Vizoso1990, Carbonell1993, Li2002}.

To compare the activity of HAT-P-11 to solar activity, we characterize the Sun's spot latitude distribution in the 1917-1985 sunspot catalog from Mt.~Wilson Observatory published by \citet{Howard1984}. We group the solar spot observations into four-year bins similar to the \kepler\ time series of HAT-P-11. On these timescales, the spot latitude distributions on both stars are often similar to Gaussians (see Figure~\ref{fig:latitude_model}), though sometimes the deviations from Gaussians are significant.

We construct a probabilistic model to describe the latitude distributions of spots on HAT-P-11 and on the Sun, following the description of the Gaussian mixture model in Section~\ref{sec:i_s}. We fit for the amplitude, mean, and variance of Gaussians representing the latitude distributions of spots in each hemisphere. We focus on fitting the shape of the latitude distribution and do not compare the total number of spots observed on the two stars to each other, since a correction for the sensitivities and biases of the different observing methods is beyond the scope of this paper.

The latitude distribution of spots of HAT-P-11 and four years of solar observations are shown in Figure~\ref{fig:latitude_model}. The four-year span of solar observations closely resembles the mean spot latitudes and standard deviation of spot latitudes that we measure for HAT-P-11. The sunspots included in Figure~\ref{fig:latitude_model} span the active maximum of solar Cycle 19, which was the solar maximum with the largest recorded number of spots since telescopic observations began \citep{Solanki2013}.

The properties of the maximum-likelihood Gaussian mixture models for HAT-P-11 and the Sun are shown in Figure~\ref{fig:sun_vs_hat11}. The circles show the mean latitudes of spots on each hemisphere of the Sun, and the standard deviations of the spot distributions. The pattern of the solar activity cycle is visible --- sunspots in the beginning of the cycle appear in small numbers at high latitudes, then large numbers near $15^\circ$, before settling back to lower numbers near the equator. The standard deviations of the spot distributions are correlated with the mean latitude --- the active latitudes are broadest at the beginning of the activity cycle when spots form at high latitudes, and the active latitudes become narrower as they approach the equator later in the activity cycle. The combined effect of the shrinking standard deviations with declining mean latitudes produces the ``wings'' in the butterfly diagram.

The distribution of spots on HAT-P-11 sits near the region of $\bar{\ell} - \sigma$ space corresponding to solar maximum. The most similar four-year bin of sunspots, which is roughly consistent with the HAT-P-11 spot distribution in terms of $\bar{\ell}$ and $\bar{\sigma}$, is the bin plotted in Figure~\ref{fig:latitude_model}. The mean active latitudes on HAT-P-11, $\ell = 16 \pm 1^\circ$, correponds to the mean latitudes of sunspots near most solar maxima.

The number of spots observed in each hemisphere is rather asymmetric, but within the range of observed asymmetries on the Sun. Hemispheric asymmetries of the solar spot distribution are often quantified by $(N-S)/(N+S)$, where $N$ is the spot area in the northern hemisphere and $S$ is the spot area in the southern hemisphere \citep{Waldmeier1971, Carbonell1993}. The maximum likelihood \stsp\ spot latitudes and radii from the entire \kepler\ mission give $(N-S)/(N+S)=0.35$. This asymmetry is within the range observed on the Sun by \citet{Howard1984} when observed in four-year bins, varying from -0.1 to 0.6.

\subsection{Radius distribution} \label{sec:radii}

\begin{figure*}
\centering
\includegraphics[scale=0.75]{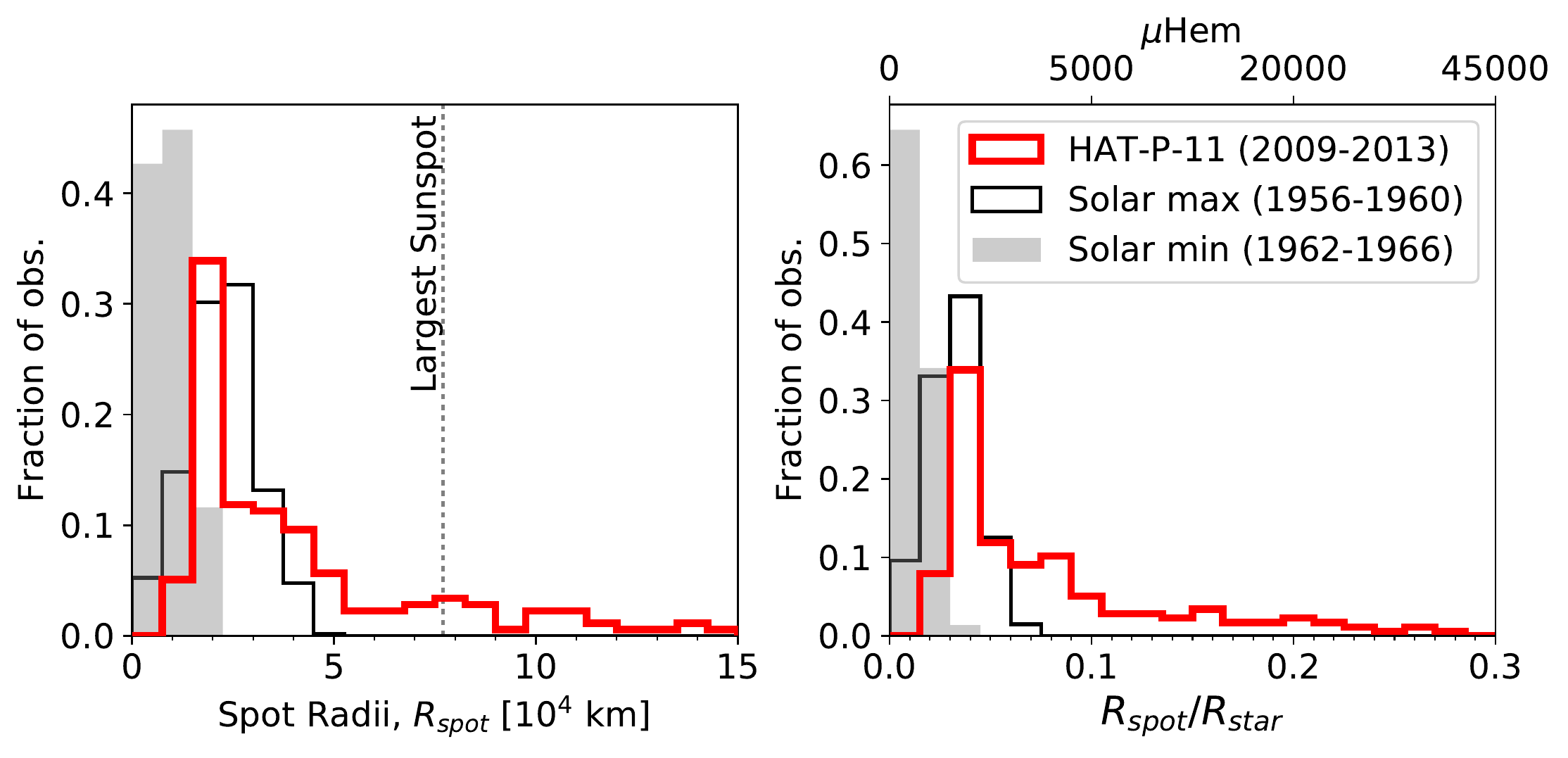}
\caption{Maximum likelihood spot radii for HAT-P-11 from \kepler\ spot occultations, and for the Sun from Mt.~Wilson Observatory \citep{Howard1984}. The spot radii are given in physical units on the left panel and in fractions of the stellar radius ($R_{spot}/R_{star}$) and millionths of the observer-facing hemisphere ($\mu$Hem) in the right panel. We adopt the radius of HAT-P-11 from \citet{bakos2010}, $R_{star} = 0.752 R_\odot$. For reference, the largest sunspot measurement we could find in the literature was 6132 $\mu$Hem = $R_{spot} / R_{\odot} = 7.7 \times 10^4$ km. We suspect that a significant fraction of HAT-P-11's spots with very large radii are in fact occultations of multiple spots.}
\label{fig:radii}
\end{figure*}

\begin{figure}
\centering
\includegraphics[scale=0.18]{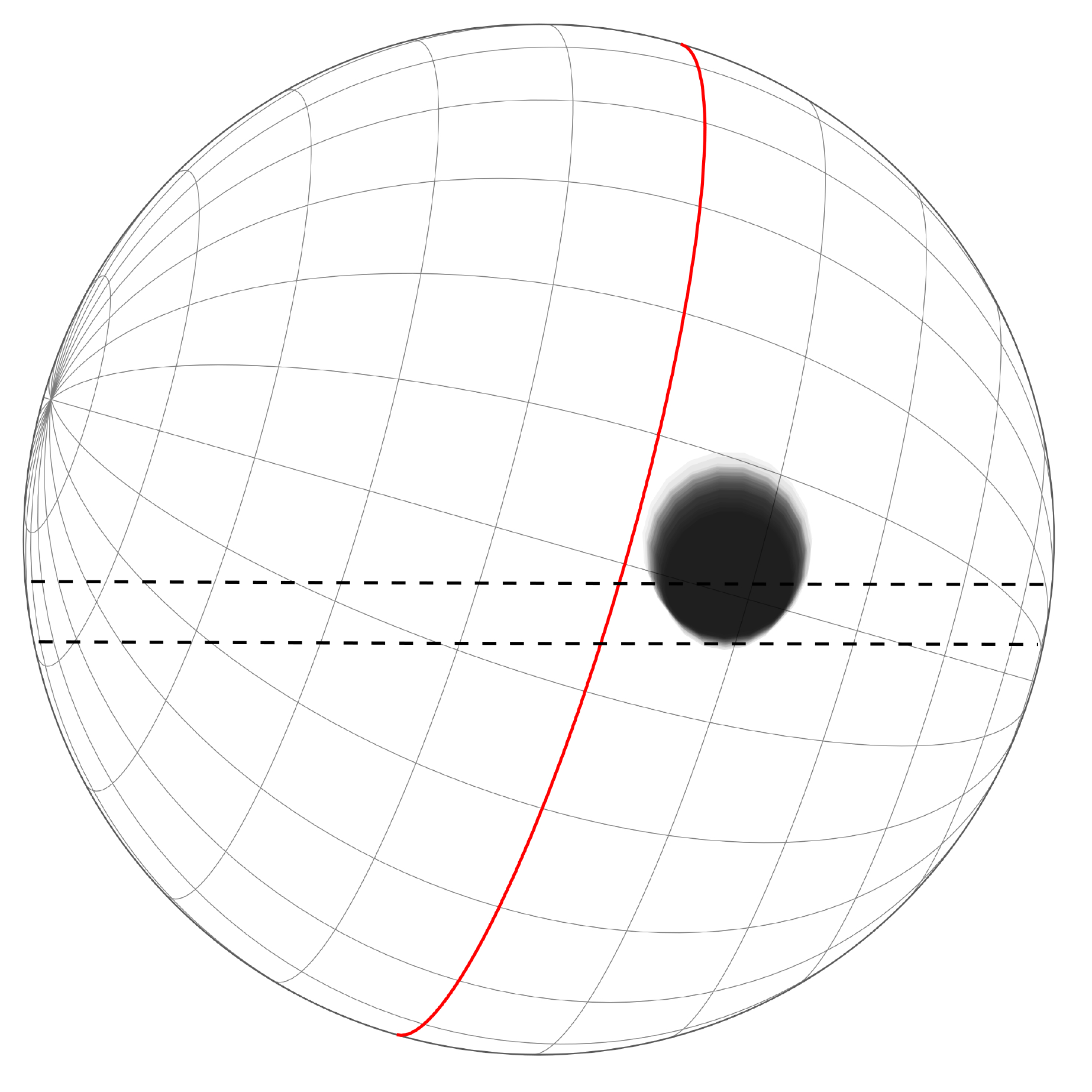}
\includegraphics[scale=0.35]{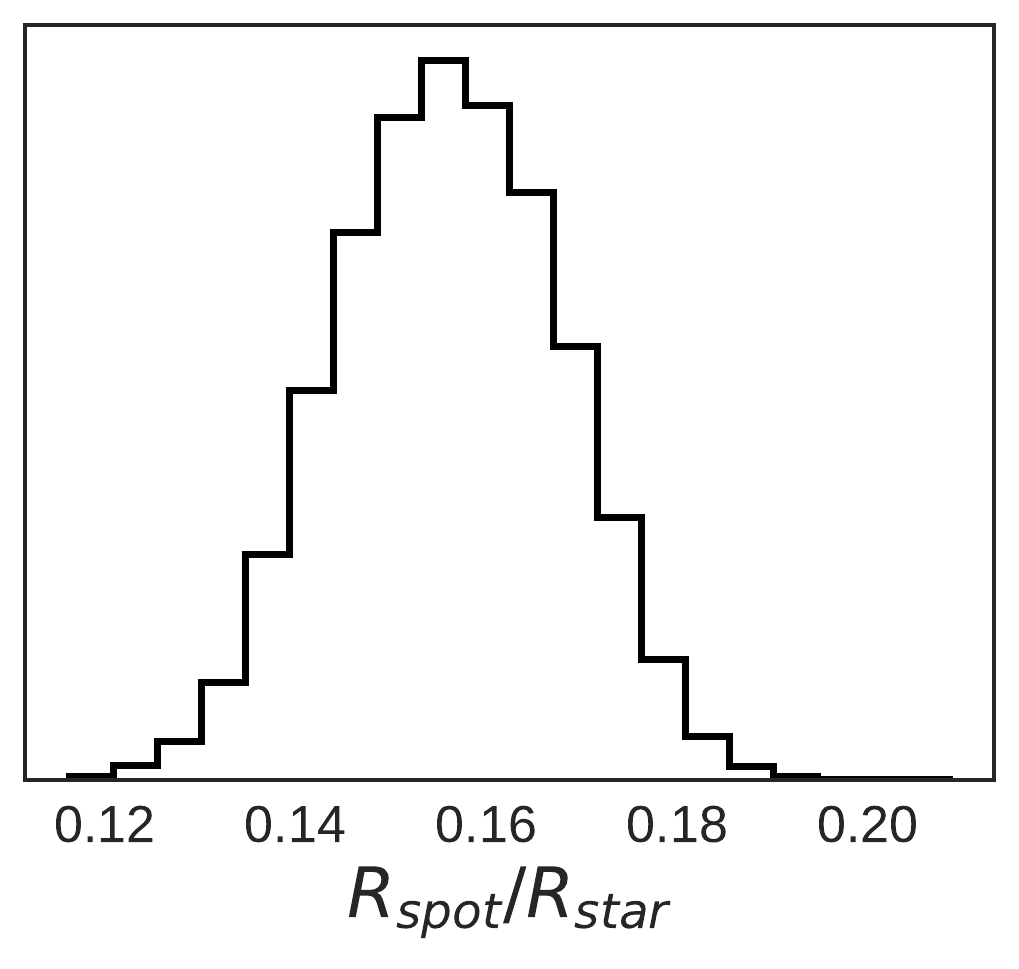}
\includegraphics[scale=0.45]{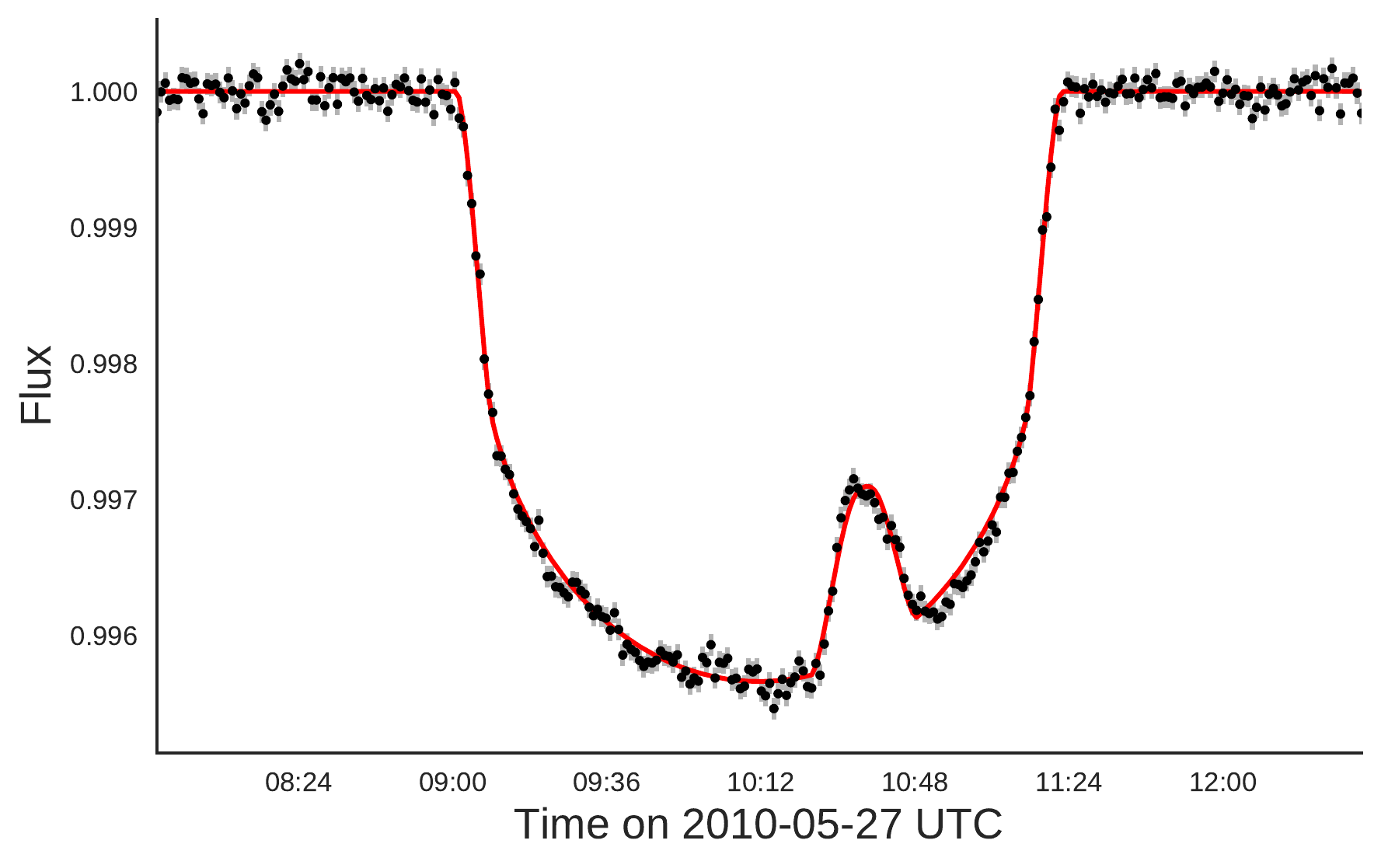}
\caption{An occultation of a particularly large spot. The maximum likelihood spot radius is $R_{spot}/R_{star} = 0.160_{-0.016}^{+0.007}$, or $84000^{+4000}_{-8000}\pm 6000$ km -- just larger than the largest recorded sunspot (77000 km, \citealt{Newton1955}). This spot is highlighted in green on the spot map in Figure~\ref{fig:map}.}
\label{fig:transit_071}
\end{figure}

We compute the physical spot radius distribution using the radius measurement of HAT-P-11 from \citet{bakos2010}, $R=0.752 \pm 0.021 R_\odot$. The spot radius distribution is shown in Figure~\ref{fig:radii}, along with sunspot radii near activity maximum and minimum. 

The spot radius distribution of HAT-P-11 closely resembles the Sun's at activity maximum for spots with radii $2\times 10^4 < R_{spot} < 5 \times 10^4$ km. 85\% of the spots on HAT-P-11 are smaller than the largest observed sunspot. The HAT-P-11 radius distribution is incomplete for small spots with $R_{spot} \lesssim 2 \times 10^4$ km, since spot occultation amplitudes of those spots are similar in scale to the noise in \kepler\ photometry. The smallest observed sunspots have radii of order $10^3$ km \citep{Solanki2003}, so it is likely that there are also small spots on HAT-P-11 below our S/N threshold.

HAT-P-11's spot distribution has a tail of spots larger than those observed on the Sun, with $R_{spot} > 5 \times 10^4$ km. From visually inspecting the individual transits, it is clear that some of the spots are larger than the largest sunspots. The largest published sunspot measurement that we encountered in the literature was recorded in 1947 by \citet{Newton1955} to have area 6132 $\mu$Hem. We can calculate the radius of a circular spot with this area in units of hemispheres (Hem) by normalizing the area of the circular spot $A_{spot} = \pi R_{spot}^2$ by the area of the observer-facing hemisphere of the star $A_{star} = 2\pi R_{star}^2$,
\begin{equation}
A_{\textrm{Hem}} = \frac{1}{2}\left(\frac{R_{spot}}{R_{star}}\right)^2\\ \\
\frac{R_{spot}}{R_{star}} = \sqrt{2 A_{\textrm{Hem}}}
\end{equation}
Therefore in the circular approximation, the largest reported sunspot had radius $R_{spot}/R_{\odot} = 0.110$ and $R_{spot} \sim 77$ Mm. We can compare that to the spot in Figure~\ref{fig:transit_071}, for example, which has $R_{spot}/R_{star} = 0.160_{-0.016}^{+0.007}$ corresponding to $R_{spot} = 84^{+4}_{-8}$ Mm --- consistent with the largest sunspot. The radius posterior distribution shown in Figure~\ref{fig:transit_071} has a single solution. However, many of these larger spots could be somewhat smaller than the maximum likelihood solution that we are reporting. The spot radius posterior distributions exhibit families of degenerate solutions in which the spot occultation fluxes can be fit equally well by a grazing spot occultation of a large spot or a more direct occultation of a smaller spot (see Section~\ref{sec:degen} for discussion of degeneracies). This could produce a systematic bias towards larger maximum-likelihood spot radii in the values that we report.

\subsection{Spotted area} \label{sec:spotted_area}

\begin{figure}
\centering
\includegraphics[scale=0.6]{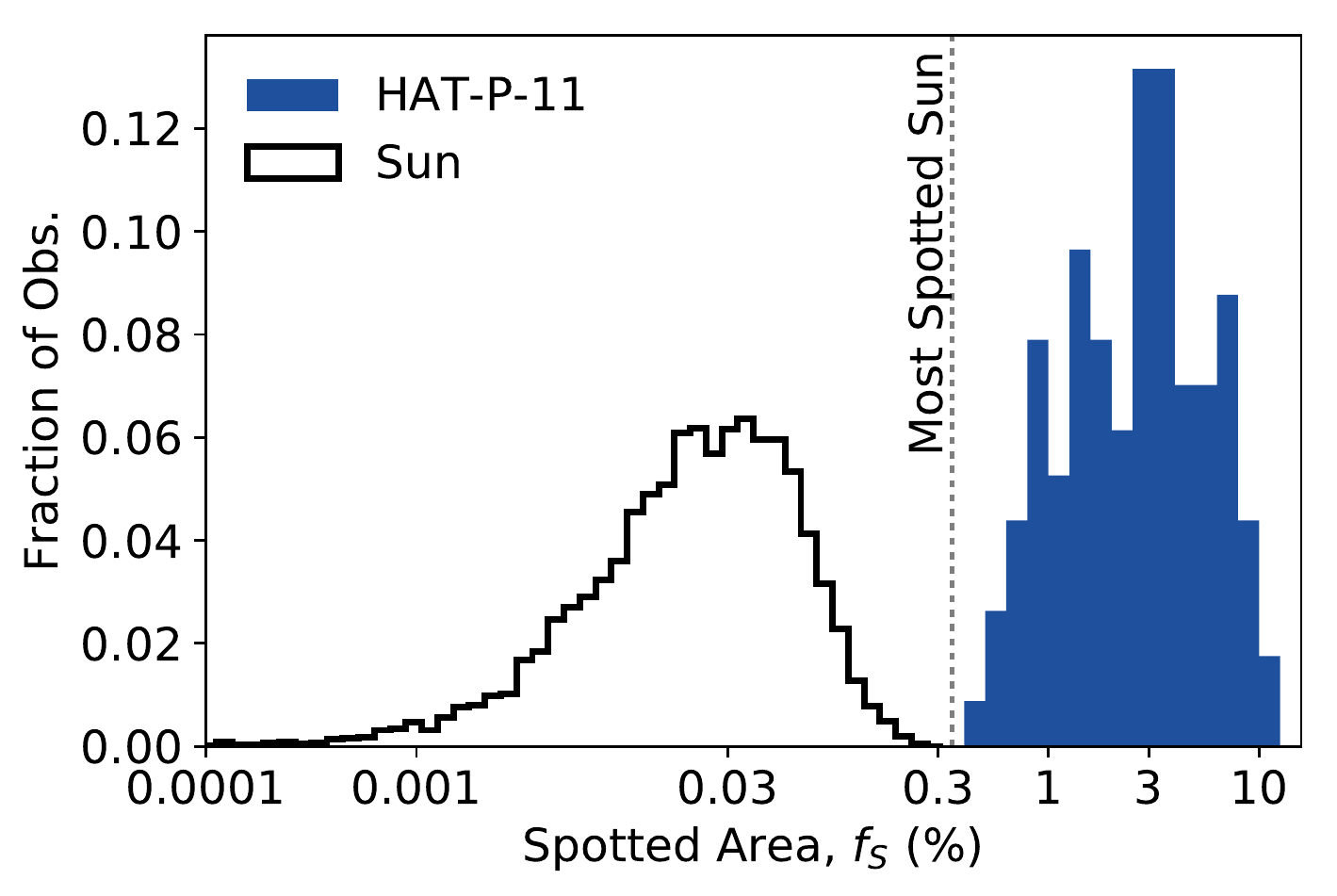}
\caption{Spot coverages of HAT-P-11 and the Sun. The HAT-P-11 spotted area is computed the area of the transit chord occulted by spots in the maximum likelihood \stsp\ fit. The spotted areas of the Sun are gathered from the Mt.~Wilson Observatory spot catalog of \citet{Howard1984}, scaled up to account for the areas of penumbra and penumbra, assuming an area ratio of $A_{pen} / A_{umb} = 4$. We have excluded spot coverages for transits that had either no significant spot occultations, or where multiple spot occultations were fit with a single unrealistically large spot, narrowing the sample down to 90 of the 205 transits observed by \kepler.}
\label{fig:spotted_area}
\end{figure}

\begin{table*}
\centering
\begin{tabular}{|lccccccp{3cm}|} 
\hline
Name & Sp. Type & $T_{eff}$ [K] & $P_{rot}$ [d] & $T_{spot}$ [K] & $f_s \, [A_{*/2}]$ & $\left < S \right>$ & Ref. \\ \hline \hline
Sun & G2V & 5777 & 24.47 & 3900-5500 & $0.0003^{+0.0006}_{-0.0001}$ & 0.17 & \citet{Howard1984, Solanki2003, Egeland2017}\\
HAT-P-11 & K4V & 4780 & 29.2 & 4500 (fixed) & $3^{+0.06}_{-0.01}$ & 0.6 & (This work, \citet{bakos2010}) \\
OU Gem  & K3V/K5V & 4925/4550 & 6.991848 & --- & $\le 0.04 - 0.35$ & 0.796 & \citet{ONeal2001, Pace2013} \\
EQ Vir  & K5Ve & 4380 & 3.96 &  $3350 \pm 115$ &$0.33 - 0.45$ & 3.68 & \citet{ONeal2001, Cincunegui2007} \\
XX Tri & K0 III & 4750 & 23.96924 & $3425 \pm 120$ & $0.31 - 0.35$ & -- & \citet{ONeal2004} \\
V833 Tau & K4V & 4500 & 1.7955 & 3175  & 0.51 & 2.460 & \citet{ONeal2004, Pace2013} \\ \hline
\end{tabular}
\caption{Comparison of HAT-P-11, the Sun, and several stars in order of increasing spot coverags ($f_S$, in units of hemispheres [$A_{*/2}$]). The spot temperature range listed for the Sun includes the typical lower limit of umbral temperatures and typical upper limit of penumbral temperatures. The spot temperature of HAT-P-11 is estimated by selecting the PHOENIX model atmosphere that most closely produces a spot contrast of $c=0.3$ in the \kepler\ bandpass \citep{Husser2013}, and should be thought of as the approximate area-weighted spot group temperature in both the umbra and penumbra. The typical solar spot temperature range spans from the coolest regions of the umbra to the hottest regions of the penumbra.}
\label{tab:fillfactors}
\end{table*}

The spot area coverage of the observable hemisphere of a star, or the spot ``filling factor'' $f_S$, has been constrained for several stars with Zeeman-Doppler imaging and molecular band modeling. Since these methods are sensitive to different spot sizes \citep{Solanki2004}, we chose to compare the spots that we detect on HAT-P-11 via photometry with starspots detected via molecular band modeling only. The molecular absorption band spot temperatures and filling factors from \citet{ONeal2001, ONeal2004} are enumerated in Table~\ref{tab:fillfactors}, with spot areas ranging from a few percent to nearly half of the stellar surface. 

We can calculate the spotted area within the transit chord at the maximum-likelihood step in the MCMC chains for each transit. This spotted area measurement is not identical to the spotted area fraction $f_S$ from \citet{ONeal2001, ONeal2004}, which measures the fractional area of spots on the entire observer-facing hemisphere of a star. However, since the transit chord of HAT-P-11 b is nearly perpendicular to the stellar equator, the planet occults most latitudes of the star at one longitude during each transit. Since we expect the distribution of starspots to be azimuthally symmetric -- i.e.~the spot distribution may change as a function of latitude but not longitude -- each transit samples the spotted area of a relatively unbiased slice of the stellar surface. Thus we use the spot coverage within the transit chord as a characteristic spot coverage on the whole star.

In Figure~\ref{fig:spotted_area}, we plot the spotted area within the transit chord for the 138 transits with significant spots modeled by \stsp, compared with the spotted area on the Sun (including both the umbra and penumbra). During the \kepler\ mission, the spotted area on HAT-P-11 varied with mean area coverage $\left< f_{s,H11} \right> = 3^{+6}_{-1} \%$, where the upper and lower error bars are the $84^\textsuperscript{th}$ and $16^\textsuperscript{th}$ percentiles, respectively. We have excluded the transits with no significant spot detections from the above reported $\left< f_{s,H11} \right>$ since the abundance of non-detections is distinct from measurements of zero spotted area; and as we discuss in Section~\ref{sec:spot_number}, the star likely always has large spots facing the observer.

The mode of the solar spot coverage from \citet{Howard1984} is $\left< f_{s,\odot} \right> = 0.0003^{+0.0006}_{-0.0001}$, $\sim$100x smaller than HAT-P-11's. Upper limits on the maximal recorded spotted area of the Sun vary depending on the observations considered, but are typically $\lesssim 0.6\%$ \citep{Balmaceda2009}.

We note that the completeness of spot detections on the Sun is nearly 100\%, whereas on HAT-P-11 we are only sensitive to large spots in the transit chord, which covers about 6\% of the observer-facing hemisphere. Therefore the spot coverage that we report for HAT-P-11 is best treated as a lower limit on the actual spot coverage. With that caveat in mind, HAT-P-11's spot coverage is most similar to the molecular band observations of OU Gem, which varies in the range $f_s \le 0.04$ to $0.35$ \citep{ONeal2001}. 

The high spot coverage of HAT-P-11 compared to the Sun is consistent with its CaII H \& K emission. The Sun's mean $S$-index during Cycle 23 was $\left<S_\odot\right> = 0.1701 \pm 0.0005$ \citep{Egeland2017}, compared to $\left<S_{H11}\right> = 0.61$ for HAT-P-11 \citep{bakos2010}. The solar $S$-index directly correlates with the area coverage by sunspots, so naturally it follows from the high $S$-index that HAT-P-11 should have a higher spot coverage. 

\citet{Shapiro2014} fit for the the relation between sunspot coverage $f_{s,\odot}$ as a function of the solar $S$-index and found:
\begin{equation}
f_{s,\odot}(S) = 0.105 - 1.315 S + 4.102 S^2. \label{eqn:naive}
\end{equation}
If we naively substitute our measured spot coverage $f_{s,H11} \sim 0.03$ for HAT-P-11 into Equation~\ref{eqn:naive}, we predict $\left<S_{H11, pred}\right> \sim 0.26$. The observed $S$-index is much larger --- evidently the activity of Sun-like stars does not scale quadratically with $S$-index in the activity regime relevant to HAT-P-11.

We now revisit the assumption made in Section~\ref{sec:norm} that the maximum flux during each \kepler\ quarter is approximately the unspotted brightness of the star. The spotted area observed on HAT-P-11 is as high as 10\% at times, so the maximum quarterly flux is unlikely to be the unspotted flux of the star. If we are underestimating the unspotted flux of the star, then we will underestimate the transit depth (see Section~\ref{sec:norm}), and therefore underestimate spot occultation amplitudes. This propagates into underestimates of spot radii and underestimates of the spotted area. This bias acts to oppose the bias towards larger spot radii and larger spotted areas discussed in Section~\ref{sec:radii}. Visual inspection of the transit light curve residuals shows that the transit depth is generally consistent with the observations, so we deem that our normalization in Section~\ref{sec:norm} is sufficient.

\subsubsection{Spotted area via flux deficit}

\begin{figure}
\centering
\includegraphics[scale=0.65]{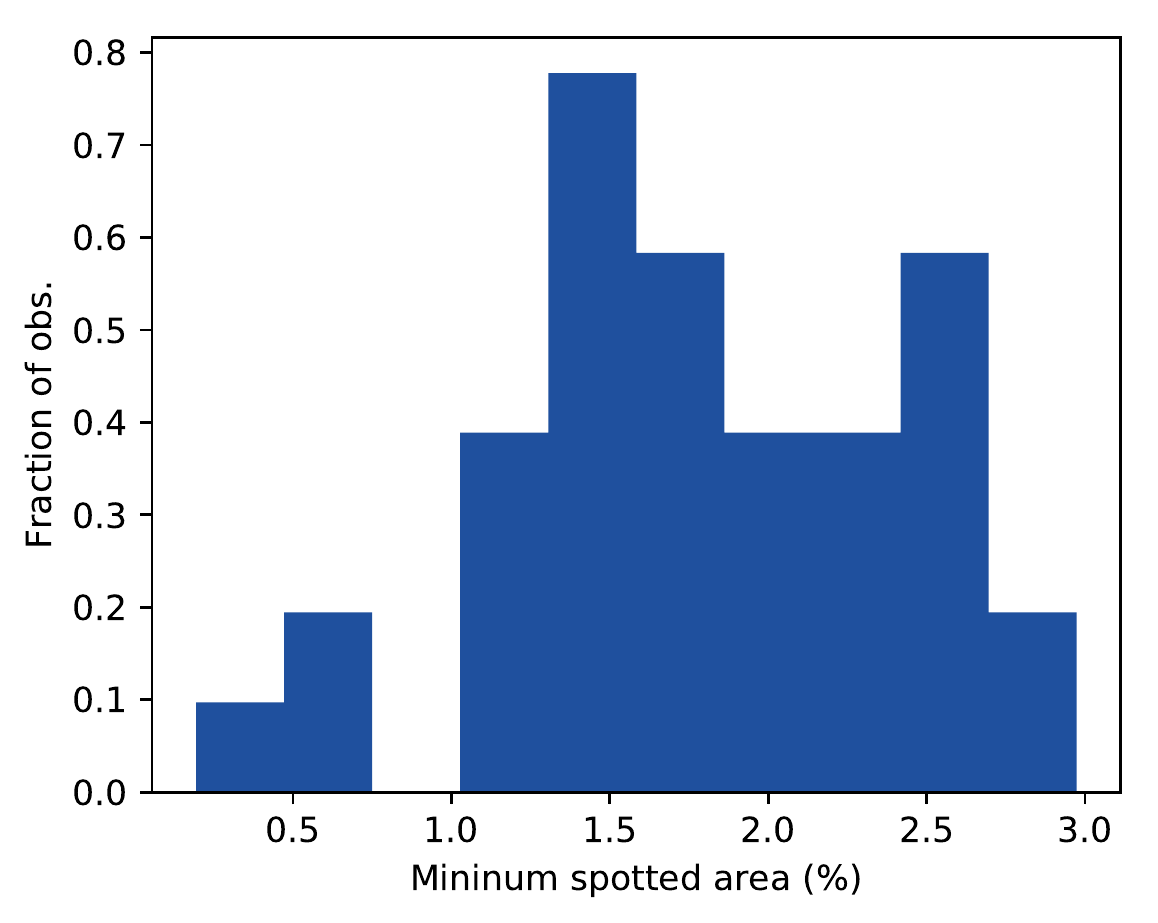}
\caption{Minimum spot coverage of HAT-P-11, independently inferred from the flux deficits in the out-of-transit portions of the \kepler\ light curve. The minimum spot coverage in the range $0.5$ to $3\%$ is consistent with the spot coverage inferred from the spotted area in the transit chord in the maximum-likelihood \stsp\ models, $f_S = 3^{+6}_{-1} \% $.}
\label{fig:flux_deficit}
\end{figure}

The rotational modulation of the out-of-transit fluxes can independently constrain the spotted area of HAT-P-11 (for an assumed spot contrast). The difference between the brightest and dimmest flux measured in each \kepler\ quarter is equivalent to the product of the fractional projected spotted area and $1-c$, where $c$ is the spot contrast as defined in Equation~\ref{eqn:contrast}. 

In practice, the estimate of the spotted area via the flux deficit is a lower limit on the total spotted area. If there were only a few small spots on the star, the flux deficit would yield the true spot covering fraction. However in the limit of many small spots distributed evenly across the star, each spot that rotates out of view will be replaced by another spot rotating into view, and thus adding more spots would not increase the flux deficit. As we argue in this section and Section~\ref{sec:spot_number}, there are a significant number of spots on the star, and it is unclear if the star is in this saturated flux deficit regime. Thus the spotted area inferred by the flux deficit should be treated as a lower limit on the spotted area.

We compute the fractional spotted area for each \kepler\ quarter from the flux deficit as follows. We mask out all transits, and convolve the fluxes with a Gaussian kernel ($\sigma=50$ fluxes). We normalize each quarter's fluxes by its smoothed maximum flux. The minimum fractional spotted area during each quarter is given by $f_S = \left( 1 - \min\textrm{(flux)} \right)/(1-c)$.

The minimum spotted area inferred by the quarterly flux deficits is shown in Figure~\ref{fig:flux_deficit}. We see that minimum spotted area ranges from 0.1\% to 3\% over the different \kepler\ quarters. This is consistent with the 0.5-10\% spotted area that we infer from modeling spots in the transit chord (Figure~\ref{fig:spotted_area}).

\subsection{Spot number} \label{sec:spot_number}

During the solar activity cycle, the number of spot groups observed on the Sun at any instant varies from near zero at activity minimum to hundreds at maximum. We measure the number of high-signal spot occultations per transit for HAT-P-11, which we can use to (1) search for evolution in the spot number over time; and (2) to compare to the number of sunspots.

The simplest measurement of the number of starspots that we can obtain from the \kepler\ photometry is the number of spots per transit for all 205 transits. Each transit is short compared to the expected spot evolution timescale (weeks) and the stellar rotation period ($29.2$ d), so each transit gives an instantaneous measurement of the number of spots within the transit chord. These spot numbers of HAT-P-11 are not directly comparable to sunspot numbers because solar observations can resolve much smaller spots than occultation photometry. It is also likely that what appear to be large spots in occultation photometry might really be groups of smaller spots at higher resolution.

We assume that the observed spot count per transit follows a Poisson distribution, and compute the likelihood of detecting the observed spot numbers for a given Poisson rate parameter $\lambda$ (with units of spots detected per transit). We allow the spot count rate to vary as a function of time, as it would throughout the solar activity cycle. We model the number of spots observed per transit with a Poisson distribution $P(\lambda)$ with a linearly varying Poisson rate parameter $\lambda(t) = \lambda_0 (t-t_0) + \lambda_1$, where $\lambda_0$ is the rate of change of the Poisson rate parameter over time (i.e.: how many more/less spots will be counted per year), and $\lambda_1$ is the rate parameter at time $t_0$. We marginalize over the hyperparameters with MCMC and find that the rate parameter slope is $\lambda_0 = 0.12 \pm 0.06$ spots per year. Since this slope is consistent with no slope, we fix $\lambda_0 = 0$ and solve only for a constant rate parameter, and find $\lambda = 0.870 \pm 0.066$ spots per transit.

The apparently constant spot number over four years could be observed for a Sun-like star with period $\sim11$ years if observed near maximum or minimum, or at any phase if the activity cycle has a long period. Spectroscopic activity index measurements over time may distinguish between these two possible cases. 

We can compute a rough estimate of the number of spots on the entire stellar surface by extrapolating from number of spots observed within each transit chord. The occulted fraction of the entire stellar surface area, $A_* = 4\pi R_*^2$, within each transit chord is $\sim 2.9\%$. If we assume there are $\lambda=0.87$ spots per transit from the analysis above, then there are $\sim 30 \pm 2$ spots like the ones detected in transit on the surface of the entire star at any given time, and about $15 \pm 1$ on the observer-facing hemisphere of the star.

We refrain from comparing the number of spots on HAT-P-11 to the common solar spot group number because solar observations are not directly analogous to the \kepler\ photometry. Ground-based observations of the Sun can observe sunspots as small as $R_{spot} = 1750$ km -- well below the smallest spots detected with high confidence on HAT-P-11 via transit photometry. However, we can compare the number of sunspots with radii as large as HAT-P-11's. The turnover in spot frequency for small spots on HAT-P-11 suggests that we are insensitive to spots with $R_{spot,min} < 2 \times 10^4$ km (see Figure~\ref{fig:radii}). We identify 898 spots larger than $2 \times 10^4$ km  observed over 10818 days on the Sun \citep{Howard1984}, which is roughly a rate of $0.08$ spots on the observer-facing hemisphere of the Sun at any instant. On HAT-P-11 we detect 130 such spots in 205 transits. The transit chord of HAT-P-11 b spans 6\% of the observer-facing hemisphere, so we expect roughly 11 spots on HAT-P-11 at any instant with radii $>2 \times 10^4$ km. Clearly there are more spots on HAT-P-11 of this size than on the Sun.

In light of the large spot number of HAT-P-11, and the sunspot-like radii of its spots, we can now interpret the spot area determined in Section~\ref{sec:spotted_area}. The spotted area on HAT-P-11 is 100x greater than solar largely due to the presence of \textit{more} spots, since the spot radii are typically quite similar to large sunspots near solar maximum (see spot radius discussion in Section~\ref{sec:radii}). 

\section{Conclusions and discussion} \label{sec:conclusion}

We have measured the properties of starspots on the active K4 dwarf HAT-P-11 from \kepler\ photometry of its transiting planet. We take advantage of the planet's well known orbital orientation to measure starspot positions during occultations by the planet. The highly misaligned orbit of the planet allows us to unambiguously resolve spot latitudes.

The spots of HAT-P-11 are similar to the Sun's in several ways. The spot contrast is consistent with the area-weighted contrast of typical sunspots, $c=0.3$ (Eqn.~\ref{eqn:contrast}). The mode of the spot radius distribution is similar to the radii of sunspots at solar maximum. The active latitudes of HAT-P-11 have the same mean latitude and standard deviation as the Sun at solar maximum. The asymmetry in the number of spots in each hemisphere is consistent with the range of values observed on the Sun. 

However, the activity of HAT-P-11 is more extreme than the Sun's. The mean spot coverage from 2009-2013 is $3^{+6}_{-1}\%$, $\sim$100x greater than the Sun's. The number of large starspots is roughly 100x greater than the number of similarly sized spots on the Sun. The $S$-index of HAT-P-11 is a factor of two greater than one would expect by extrapolating from the spot coverage--$S$-index relation observed on the Sun.

The similarities between the spot distributions on the Sun and HAT-P-11 are interesting in the context of dynamo theory \citep[e.g.][]{Charbonneau2010}. This K4 star is not fully convective, and therefore is expected to have a tachocline like the Sun. Perhaps the $\alpha\Omega$ dynamo is operating within HAT-P-11 as it does in the Sun. It seems that a $0.8M_\odot$ star with a near-solar rotation rate produces starspots in strikingly similar active latitudes, with more large spots. The theoretical prescriptions for magnetic flux emergence developed for the Sun may therefore be applicable out to spectral type K4 \citep[e.g.][]{Cheung2014}.

Precision spot occultation analysis made possible by \kepler\ could potentially be reproduced with photometry from NASA's TESS mission for HAT-P-11 in particular, and for active planet-host stars in general \citep{Ricker2014}. However, the one-minute cadence photometry was critical for resolving the spot occultation features of HAT-P-11, and time resolution directly translates to latitude resolution for highly misaligned systems. The TESS mission's planned two-minute cadence is likely sufficient to detect spot occultations in systems like HAT-P-11, though shorter cadence ground-based photometry would be preferred.

\subsection{Future Work}

HAT-P-11 is exceptionally bright ($V=9.47$), which makes ground-based observations of spot occultations with amplitudes on the order of 0.1\% feasible. In particular, we plan to collect transit photometry with the holographic diffuser and the ARCTIC imager on the ARC 3.5 m Telescope at the Apache Point Observatory (APO) (Stef\'{a}nsson et al.~2017, submitted). If HAT-P-11 exhibits evolution in the spot latitude distribution like the Sun does, we may be able to observe changes in the mean spot position as the activity cycle progresses. Observing spot occultations from the ground is advantageous because the latitude resolution is linked to the time resolution of the photometry, which can be minimized with large aperture telescopes and thus shorter exposure times compared to \kepler\ or TESS.

The phase of the activity cycle of HAT-P-11 can be constrained over several years by analyzing long-term spectroscopy of the $S$-index. In Morris et al.~(2017, in prep), we constrain the period and amplitude of the activity cycle of HAT-P-11 using archival high resolution spectroscopy of the star, in combination with recent high resolution spectra obtained at APO.

The constraints on the spot coverage of HAT-P-11 from the \kepler\ photometry are complementary to spectroscopic constraints from molecular band modeling. The spot coverage reported here could be independently measured by modeling absorption by TiO and OH in starspots \citep{ONeal2001, ONeal2004}. 

In this work we limited ourselves to studying only the spot occultations in transit to make direct comparisons between spots on HAT-P-11 and sunspots. Simultaneous modeling of the out-of-transit fluxes would provide complementary constraints on the total spot coverage on HAT-P-11. 

\acknowledgments

We acknowledge support from NSF grant AST-1312453. We thank Eric Agol, John Lurie, Dan Foreman-Mackey and Charli Sakari for constructive conversations during the development of this work.

Some of the data presented in this paper were obtained from the Mikulski Archive for Space Telescopes (MAST). STScI is operated by the Association of Universities for Research in Astronomy, Inc., under NASA contract NAS5-26555. Support for MAST for non-HST data is provided by the NASA Office of Space Science via grant NNX09AF08G and by other grants and contracts. This research has made use of NASA's Astrophysics Data System. This work used the Extreme Science and Engineering Discovery Environment (XSEDE), which is supported by National Science Foundation grant number ACI-1548562. This research was done using resources provided by the Open Science Grid [1,2], which is supported by the National Science Foundation award 1148698, and the U.S. Department of Energy's Office of Science. This work was facilitated though the use of advanced computational, storage, and networking infrastructure provided by the Hyak supercomputer system and funded by the STF at the University of Washington.

\facility{Kepler}

\software{STSP \citep{Hebb2017}, \texttt{ipython} \citep{ipython}, \texttt{numpy} \citep{VanDerWalt2011}, \texttt{scipy} \citep{scipy},  \texttt{matplotlib} \citep{matplotlib}, \texttt{astropy} \citep{Astropy2013}, \texttt{batman} \citep{Kreidberg2015}, \texttt{gatspy} \citep{gatspy}}

\appendix

\section{Spot contrast} \label{sec:appendix_contrast}

We make some simplifying assumptions to derive constraints on the spot contrast from the amplitudes of spot occultations, and we generalize the formalism later. We will at first calculate the flux only for spot-planet orientations where the planet completely occults the spot or the spot completely encompasses the planet. By ignoring grazing spot occultations, we will calculate maximum spot-occultation amplitudes, since grazing spot occultations yield smaller amplitudes than complete occultations. We also ignore stellar limb darkening.

The flux lost during the transit of a planet with radius $R_p$ across an unspotted star with radius $R_\star$ without limb darkening is
\begin{eqnarray}
\delta_{unspotted} = \frac{\Delta F}{F_\star} = \frac{I_\star \pi R_p^2}{I_\star \pi R_\star^2} = \frac{R_p^2}{R_\star^2}
\end{eqnarray}
where $I_\star$ is the mean surface intensity of the stellar disk per unit area. 
\begin{figure}
\centering
\includegraphics[scale=0.4]{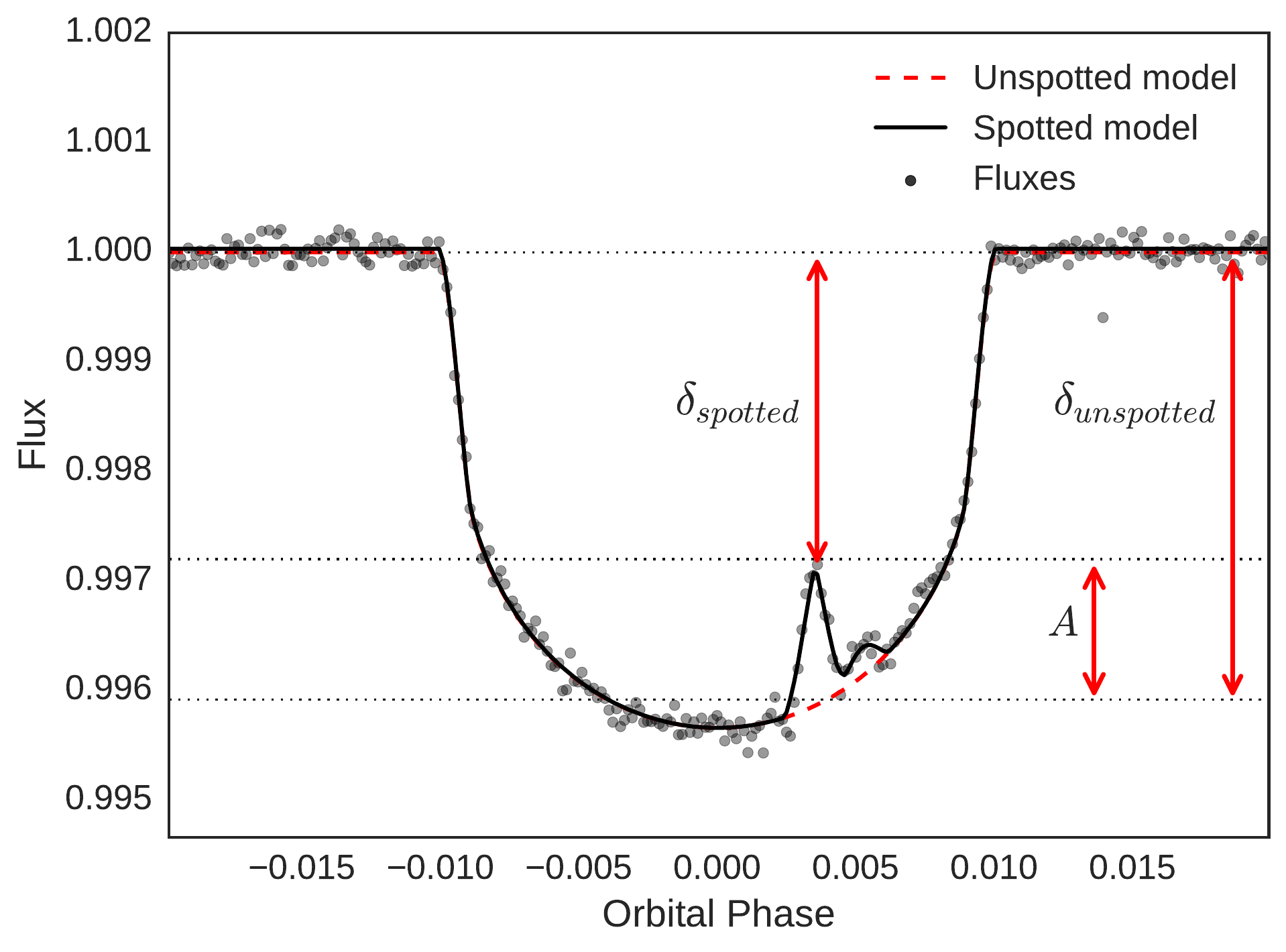}
\caption{Schematic for parameter definitions in Section~\ref{sec:contrast}, plotted on the transit of HAT-P-11 b on August 20, 2011 UT. The spot occultation amplitude $A$ is the difference between the flux lost during a transit with no spot occultations and the flux lost during a transit with spot occultations, $A = \delta_{unspotted} - \delta_{spotted}$. Note that in this terminology the ``depth'' $\delta(\mu) = \Delta F(\mu)/F$ is a function of the sky-projected distance between the planet and the star $\mu$, or equivalently time or orbital phase, for a star with limb-darkening.}
\label{fig:contrast_schematic}
\end{figure}

We measure the amplitude of brightening during a spot occultation $A$, see Figure~\ref{fig:contrast_schematic} for a schematic representation. During an occultation of a starspot, the appropriate formula for the observed flux depends on the size of the spot $R_{sp}$ relative to the size of the planet $R_{p}$. If the spot with radius larger than or equal to the radius of the planet $R_{sp} \ge R_p$ and the spot has contrast $c$, the amplitude of the difference in flux between a transit of an unspotted and a spotted star is
\begin{eqnarray}
 A &=& \left. \delta_{unspotted} - \delta_{spotted} \right|_{R_{sp} > R_p} \\
 &=& \frac{I_\star \pi R_p^2}{I_\star \pi R_\star^2} - \frac{((1 - c)I_\star) \pi R_p^2}{ I_\star \pi R_\star^2}\\
A/\delta_{unspotted} &=&  c. \label{eqn:bigspot}
\end{eqnarray}
For a spot smaller than the planet, the difference between the spotted and unspotted flux is
\begin{eqnarray}
 A &=& \left. \delta_{unspotted} - \delta_{spotted} \right|_{R_{sp} < R_p} \\
 &=& \frac{I_\star \pi R_p^2}{I_\star \pi R_\star^2} - \frac{I_\star \pi R_p^2 - c I_\star \pi R_p^2 R_{sp}^2/R_p^2}{ I_\star \pi R_\star^2}\\
 A/\delta_{unspotted} &=&  c \left(\frac{R_{sp}}{R_p}\right)^2
 \label{eqn:littlespot}
\end{eqnarray}

In the small planet limit where $R_p/R_\star \ll 1$, the stellar limb darkening could be defined, for example, with a quadratic law $I(\mu)/I_0 = 1 - u_1(1-\mu) - u_2(1-\mu)^2$, and the instantaneous unspotted depth becomes 
\begin{equation}
\delta_{unspotted}(\mu) =  \frac{R_p^2}{R_\star^2} \left[ \frac{1 - u_1(1-\mu) - u_2(1-\mu)^2}{1 - \frac{1}{3}u_1 - \frac{1}{6}u_2} \right]
\end{equation}
where $\mu$ is the sky-projected distance between the planet and the star. Equations \ref{eqn:bigspot} and \ref{eqn:littlespot} above can be generalized for stars with limb-darkening by replacing $\delta_{unspotted} \rightarrow \delta_{unspotted}(\mu)$.

% \bibliography{bibliography}

\end{document}